\documentclass[12pt,psfig,notitlepage]{article}
\usepackage{graphicx,setspace,float,rotating,longtable,amsmath,verbatim,amssymb,epsfig,color,lscape,subfigure,url}
\usepackage{graphics,caption}
\usepackage{verbatim,color,amssymb,epsfig,setspace}
\usepackage{amsmath,amsthm,amssymb,enumerate}
\usepackage{ctable}
\usepackage{epstopdf}
\usepackage{apacite}
\usepackage[english]{babel}
\usepackage{setspace,amsthm}
\newtheorem{thm}{Theorem}[section]

\theoremstyle{definition}
\newtheorem{definition}[thm]{Definition}

\theoremstyle{remark}

\def\bse{\begin{eqnarray*}}
\def\ese{\end{eqnarray*}}
\def\be{\begin{eqnarray}}
\def\ee{\end{eqnarray}}
\def\boxit#1{\vbox{\hrule\hbox{\vrule\kern6pt\vbox{\kern6pt#1\kern6pt}\kern6pt\vrule}\hrule}}

\begin{document}

\title{Robust Hybrid Learning for Estimating Personalized Dynamic Treatment Regimens }
\author{Ying Liu, Yuanjia Wang,  Michael R. Kosorok, Yingqi Zhao, Donglin Zeng\footnote{\baselineskip=12pt Ying Liu is currently an Assistant Professor (yiliu@mcw.edu) at Medical College of Wisconsin and this work was submitted when she was a Ph.D. student at Columbia University,, Yuanjia Wang is an Associate
  Professor (yuanjia.wang{\it @}columbia.edu), Department of
  Biostatistics, Mailman School of Public Health,  Columbia University, New York, NY 10032. Michael R. Kosorok is Kenan Distinguished Professor, Department of Biostatistics, University of North Carolina at Chapel Hill (kosorok{\it @}bios.unc.edu). Yingqi Zhao is Assistant Professor, Department of Biostatistics and Medical Informatics, University of Wisconsin at Madison (yqzhao{\it @}biostat.wisc.edu). Donglin Zeng is Professor, Department of Biostatistics, University of North Carolina at Chapel Hill (dzeng{\it @}email.unc.edu). This work is supported by NIH grants NS073671, P01CA142538, National Center for Research Resources grant UL1 RR025747.}}
\date{September 17, 2014}

\maketitle

\begin{abstract}\setlength{\baselineskip}{12pt}
Dynamic treatment regimens (DTRs) are sequential decision rules tailored at each stage by potentially time-varying patient features and intermediate outcomes observed in previous stages. The complexity, patient heterogeneity and chronicity of many diseases and disorders call for learning optimal DTRs which best dynamically tailor treatment to each individual's response over time. Proliferation of personalized data (e.g., genetic and imaging data) provides opportunities for deep tailoring as well as new challenges for statistical methodology. In this work, we propose a robust hybrid approach referred as Augmented Multistage Outcome-Weighted Learning (AMOL) to integrate outcome-weighted learning and Q-learning to identify optimal DTRs from the Sequential Multiple Assignment Randomization Trials (SMARTs). We generalize outcome weighted learning (O-learning; Zhao et al.~2012) to allow for  negative outcomes; we propose methods to reduce variability of weights in O-learning to achieve numeric stability and higher efficiency; finally, for multiple-stage SMART studies, we introduce doubly robust augmentation to machine learning based O-learning to improve efficiency by drawing information from regression model-based Q-learning at each stage. The proposed AMOL remains valid even if the Q-learning model is misspecified. We establish the theoretical properties of AMOL, including the consistency of the estimated rules and the rates of convergence to the optimal value function.  The comparative advantage of AMOL over existing methods is demonstrated in extensive simulation studies and applications to two SMART data sets: a two-stage trial for attention deficit and hyperactive disorder (ADHD) and the STAR*D trial for major depressive disorder (MDD).
\end{abstract}


\section{Introduction}

Advances in technology in  data collection is revolutionizing medical research by providing personal data (e.g., clinical assessments, genomic data, electronic health records) for physicians and clinical researchers to meet the promise of individualized treatment and health care. Proliferation of these data provides new opportunities to deeply tailor treatment for each subject while at the same time posing new challenges for statistical methodology. The goal of this work is to develop new and powerful methods to discover optimal  dynamic treatment regimes (DTRs, \citeA{Lavori2000})   for personalized medicine in this era of abundant data.

The complexity, patient heterogeneity and chronicity of many diseases and disorders call for designing treatment regimes that can dynamically adapt to an individual's response over time and  improve long term clinical benefits. Sequential treatments, a sequence of interventions in which the treatment decisions adapt to the time-varying clinical status of a patient, are useful in treating many complex chronic disorders such as substance dependence, attention deficit hyperactivity disorder (ADHD), autism spectrum disorder (ASD), major depressive disorder (MDD), and schizophrenia \cite{schneider2003,rush2004,rush2006,naltr,preg}. 
DTRs operationalize this process of medical decision-making and can be used to inform clinical practice. DTRs are sequential decision rules, tailored at each stage by time-varying patient features and intermediate outcomes in the previous stages.  They are also known as adaptive treatment strategies \cite{Lavori2000}, multi-stage treatment strategies \cite{thall2002, thall2005} and treatment policies \cite{Lunceford2002, wahed2004, wahed2006}.

Sequential Multiple Assignment Randomized Trials (SMARTs) were proposed  \cite{Murphy2005} to best construct DTRs that offer causal interpretation through randomization at each critical decision point.
Methods to identify optimal DTRs from SMARTs data  have recently received attention in the statistical community \cite{lavori2004,murphy2003,robins2004,moodie2007, robust2012,robust2013,zhao2009,zhao2012,zhao2013}.
See \citeA{chakraborty2013statistical} for a detailed review of the current literature.  Parametric methods based on G-computation and Monte Carlo simulation \cite{lavori2004} and  Bayesian approaches \cite{wathen2008,arjas2000,arjas2010} were proposed. These methods are subject to model misspecification and computationally intensive to implement, especially in the presence of high dimensional tailoring variables. Instead of modeling the full data generation scheme, semi-parametric methods have been developed to focus on the contrast of conditional mean outcomes.
They include methods using the ``regret function",  the expected difference in the outcomes that would have been observed if patients had taken the optimal treatment in the current stage \citeA{murphy2003}, which is also a special case of G-estimation proposed by  \cite{robins2004, moodie2007}. Some other approaches include regret regression methods  \cite{almirall2010}, marginal structural models and structural nested mean models \cite{chakraborty2013statistical}. \citeA{robust2012} and \citeA{robust2013} developed a method based on augmented inverse probability weighting for a single or multiple stage model. However, all these methods are based on either parametric or semiparametric models which may lose prediction accuracy in the presence of model misspecification and a large number of tailoring variables, especially for small or moderate sample sizes in real trials.

Machine learning methods originating from computer science literature have also been recently introduced to identify optimal treatment regimes using data collected in SMARTs. These methods are particularly suitable for dealing with a large number of subject-specific variables. For example, Q-learning first proposed in \citeA{watkins1989} was implemented to analyze SMART data by \citeA{murphy2007,zhao2009}. It is a regression-based method to identify the optimal multi-stage decision rules, where the optimal treatment at each stage is discovered by a backward induction to maximize the estimated Q-function (``Q" stands for ``quality of action''). 
The regression based Q-learning can suffer from incorrect model assumptions.  The approach selects the optimal treatment by modeling the Q-function and its contrasts which are not explicitly  related to the optimization of the objective function. The mismatch between maximizing the Q-function and the value function potentially leads to incorrect regimes due to overfitting the regression model.
To remedy these limitations,
\citeA{zhao2012} proposed an outcome weighted learning (O-learning) which chooses the treatment rules by directly optimizing the expected clinical outcome for single stage trials. The resulting optimal treatment regimen can take any unconstrained  functional form. Their simulation studies demonstrate that O-learning outperforms Q-learning especially in small sample size settings with a large number of tailoring variables. More recently, \citeA{zhao2014} generalized the O-learning to deal with multiple stage problem by a backward iterative method.
A direct comparison with \citeA{robust2013} was performed in \citeA{zhao2014} to demonstrate greater value function achieved by O-learning based methods.

In this paper, we propose a hybrid approach to integrate O-learning and regression models such as Q-learning, namely, Augmented Multi-stage Outcome-weighted Learning (AMOL), to identify the optimal DTRs from SMARTs.  We introduce doubly robust augmentation to machine learning based O-learning to improve efficiency by drawing information from regression model-based Q-learning at each stage. Thus we take advantage of the robustness from the former and the imputation ability of the latter.
Compared to \citeA{zhao2012} and \citeA{zhao2014}, there are three new contributions in this work. First, for single-stage trials, AMOL generalizes the original O-learning in \citeA{zhao2012} to allow for negative outcome values instead of adding an arbitrarily large constant as in \citeA{zhao2012} which may lead to numeric instability. This feature is practically appealing when there are both positive and negative outcomes observed in a clinical study (e.g., rate of change of clinical symptoms).  Secondly, by using residuals from a regression on variables other than the treatment assignment as outcome values, AMOL is able to reduce the variability of weights in O-learning to achieve numeric stability and efficiency gain.
Thirdly, and most importantly, for multiple-stage trials, AMOL estimates optimal DTRs via a backward induction procedure that makes non-trivial extension of \citeA{zhao2014} to boost efficiency through augmentation and integration with Q-learning.

Specifically, at each stage, AMOL uses robustly weighted O-learning for estimating the optimal DTRs, where the weights are based on the observed outcomes and conditional expectations for subjects who follow the optimal treatment rules in future stages, while for those who do not follow optimal rules in future stages, the weights are imputed from regression models obtained in Q-learning. Therefore, AMOL, as a hybrid approach, simultaneously takes advantage of the robustness of nonparametric O-learning and makes use of the model-based Q-learning utilizing data from all subjects. Moreover, AMOL still yields the correct optimal DTRs even if the regression models assumed in the Q-learning are incorrect, thus maintaining the robustness of O-learning. Similar to the double robust framework in missing data literature, AMOL aims to strike a balance between robustness and efficiency under  doubly robust estimation. However, important distinctions between the current work and the existing doubly robust estimators include the dynamic feature of the problem and the objective function used to maximize clinical benefits.

The rest of this paper is organized as follows. In Section 2, we review some concepts for DTR, Q-learning and O-learning, and introduce  AMOL through doubly robust augmentation. Section 3 presents theoretical results to justify the optimality of AMOL. Section 4 shows the results of extensive simulation studies to examine the performance of AMOL compared to Q-learning and O-learning. In Section 5, we present data analysis results based on  SMART data for an ADHD trial \cite{pelham2008}. We present another real data example on the STAR*D trial \cite{rush2004} for MDD in the Online Supplement. Lastly, we conclude with a few remarks in Section 6.

\section{Methodologies}
\subsection{Dynamic Treatment Regimes (DTRs) and O-Learning}
We start by introducing notation for a $K$-stage DTR. For $k=1,2,\dots,K$, denote $X_k$ as the observed subject-specific  tailoring variables collected just prior to the treatment assignment at stage $k$. Denote $A_k$ as the treatment assignment taking values $\{-1,1\}$, and denote as $R_k$ the clinical outcome (also known as the ``reward'') post the $k$th stage treatment. Larger rewards may correspond to better functioning or fewer symptoms depending on the clinical settings. A DTR is a sequence of decision functions, ${\cal D}=({\cal D}_1, {\cal D}_2,\dots,{\cal D}_K)$, where ${\cal D}_k$ maps the domain of patient health history information $H_k=(X_1,A_1,\dots,A_{k-1},X_k)$ to the treatment choices in $\{-1,1\}$. Let $\pi_k(a,h)$ denote the treatment assignment probability, $\textrm{Pr}(A_k=a|H_k=h)$. In a SMART design, it is the randomization probability for each treatment group at each stage specified by design and thus known to the investigator. Our goal is to maximize the total reward from all stages, i.e., $\sum_{k=1}^{K}R_k$.
To evaluate the causal effect of a DTR, potential outcome framework in causal inference literature is used. The potential outcome in our context is defined as the outcome of a subject had he/she followed a particular treatment regimen, possibly different from the observed regimen.
Several assumptions are required to infer causal effects of a DTR using data from a SMART, including the standard stable unit treatment value assumption and the no unmeasured confounders assumption \cite{murphy2001,moodie2007}. Furthermore, we need the following positivity assumption: for $k=1,...,K$ and any $a_k\in \{-1,1\}$ and $h_k$ in the support of $H_k$, $\pi_k(a_k,h_k)=P(A_k=a_k|H_k=h_k) \in (c_0,c_1)$, where $0<c_0<c_1<1$ are two constants.
Note that the no unmeasured confounders assumption is satisfied for SMART studies due to the virtue of sequential randomization, while the positivity assumption requires that each DTR has a positive chance of being observed.

Let $P$ denote the probability measure generated by $(X_1,A_1,R_1,\dots,X_K,A_K,R_K)$. Let $P_{\cal D}$ denote the conditional probability measure generated by random variables
$(X_1, A_1, R_1,\dots,$ $X_K, A_K, R_K)$ given that regimen ${\cal D}$ is used to assign treatment, i.e., $A_k={\cal D}_k(H_k)$. Denote $E_{\cal D}$ as the conditional expectation with respect to the distribution $P_{\cal D}$.  The value function associated with ${\cal D}$ is defined as the expected reward given that the treatment assignments follow regimen ${\cal D}$ \cite{qian2011}, that is,
${\cal V}({\cal D})=E_{\cal D}\left[\sum_{k=1}^{K}R_k\right]=\int \sum_{k=1}^{K}R_k \,\mathrm{d} P_{\cal D}. $
Under the positivity assumption, it is easy to show that $P_{\cal D}$ is dominated by $P$ and
\begin{equation}
{\cal V}({\cal D})=E\left[\frac{\prod_{k=1}^{K}I(A_k={\cal D}_k(H_k))(\sum\limits_{k=1}^KR_k)}{\prod\limits_{k=1}^{K}\pi_k(A_k,H_k)}\right].
\label{valuefunction}
\end{equation}
Hence, the goal is to find the optimal treatment rule ${\cal D}^*$ that maximizes the value function, that is, ${\cal D}^*=\textrm{argmax}_{{\cal D}} {\cal V}({\cal D}).$ {Without confusion, sometimes we write ${\cal V}({\cal D})={\cal V}(f_1,...,f_K)$ if ${\cal D}_k$ is given as the sign of $f_k$.}

Recently, O-learning (Zhao et al., 2012; 2014) is proposed to estimate the optimal treatment regimes. Specifically, for a single stage study,  O-learning reformulates the value optimization problem as a weighted classification problem where the weights are the observed rewards, and a weighted support vector machine was proposed to estimate the optimal treatment decision \cite{zhao2012}. Based on $n$ observations $(H_i,A_i,R_i)$ with randomization probability $\pi_i=\pi(A_i,H_i)$,  O-learning minimizes a surrogate hinge loss with $L_2$-regularization:
\begin{equation}
\min_{f}n^{-1}\sum_{i=1}^n(1-A_if(H_i))^+\frac{R_i}{\pi_i}+\lambda_n\|f\|^2.
\label{iowl}
\end{equation}
Here, $f(\cdot)$ is a parametric or nonparametric decision function, $\Vert f\Vert$ denotes some norm or semi-norm of $f$ such as the norm in a reproducing kernel Hilbert space (RKHS), and the corresponding treatment rule is ${\cal D}(H_i)=\textrm{sign}(f(H_i))$. For a multiple-stage study, Zhao et al.~(2014) proposed a backward induction to implement O-learning where at stage $k$, they only used the subjects who followed the  estimated optimal treatment regimes after stage $k$ in the optimization algorithm. That is, the optimal rule ${\cal D}_k(H_k)=\textrm{sign}(f_k(H_k))$
solves
$$
\min_{f}n^{-1}\sum_{i=1}^n(1-A_{ik}f_k(H_{ik}))^+\frac{R_{ik}+\cdots +R_{iK}}{\pi_{ik}}\frac{I(A_{i,k+1}=\widehat{\cal D}_{k+1}(H_{i,k+1}),\cdots, A_{iK}=\widehat{\cal D}_K(H_{iK}))}{\prod_{j>k}\pi_{ij}}$$
$$+\lambda_n\|f_k\|^2,$$
where $(A_{ik}, H_{ik}, R_{ik})$ denotes the observation at stage $k$, $\pi_{ij}=\pi_j(A_{ij}, H_{ij})$, and $\widehat {\cal D}_j$ is the estimated optimal rule at stage $j$ from the backward learning. However, as discussed before, only using the subjects who followed the optimal regimes in future stages may result in much information loss when $K$ is not small. Furthermore, their method suggests to subtract a constant from $R_{ik}$'s to ensure a positive weight in the algorithm, where the choice of constant  may be arbitrary.

\subsection{General Framework to Improve O-Learning}

In this section, we describe theoretical motivation and general ideas to improve O-learning.
First consider a single-stage trial with data $(A_i,H_i,R_i), i=1,...,n$. Detailed implementation of the improved O-learning for a multiple-stage study will be given in the next section.

The key idea to improve O-learnings is to replace weights $R_i$ in (\ref{iowl}) by surrogate weights   $\widetilde R_i$ so that under certain regularity conditions, we achieve two desirable properties: (a) the treatment rule minimizing
\begin{equation}
\min_{f}n^{-1}\sum_{i=1}^n(1-A_if(H_i))^+\frac{\widetilde R_i}{\pi_i}+\lambda_n\|f\|^2
\label{iowlnew}
\end{equation}
is still asymptotically consistent; and (b) the derived rule has smaller variability {compared to using $R_i$ as weights}.
Property (a) ensures that using new weights $\widetilde R_i$ still yields a correct optimal treatment regime, while property (b) implies that the new rule is estimated more accurately than the one given by O-learning.
To achieve these goals, first observe that as shown in Zhao et al.~(2012), the optimal rule minimizing (\ref{iowlnew}), assuming that $\lambda_n$ vanishes asymptotically, has the same sign as $E[\widetilde R|A=1, H]-E[\widetilde R|A=-1, H]$. Therefore, {property (a) is asymptotically equivalent to requiring}
\begin{equation}\textrm{sign}\left\{E[R|A=1, H]-E[R|A=-1, H]\right\}=
\textrm{sign}\left\{E[\widetilde R|A=1, H]-E[\widetilde R|A=-1, H]\right\}.\label{optima}
\end{equation}
For (b), to examine accuracy of the estimated rule, we assume $f(h)$ to be a parametric function $\theta^T\phi(h)$ where $\phi(h)$ is a vector of some pre-defined basis functions. Thus, under the same regularity conditions in \cite{koo2008bahadur} and following their proofs, we can show that the asymptotic variance of the estimator for $\theta$ is
$$
E\left[\widetilde R \delta(1-A f^*(H))\phi(H)\phi(H)^T/\pi(A,H)\right]^{-1}
E\left[\widetilde R^2 I(1-Af^*(H)>0)\phi(H)\phi(H)^T/\pi(A,H)^2\right] $$
\begin{equation}
\times E\left[\widetilde R \delta(1-A f^*(H))\phi(H)\phi(H)^T/\pi(A,H)\right]^{-1},
\label{optimb}
\end{equation}
where $\delta(\cdot)$ denotes the Dirac function and $f^*$ is the unique minimizer for $E[\tilde R(1-Af(H))^+].$ Consequently, from (\ref{optima}) and (\ref{optimb}), we conclude that if $\widetilde R$ satisfies
$$E[\widetilde R|A,H]=E[R|A,H], \ \ Var(\widetilde R|A,H)\le Var(R|A,H),$$
then since
$$E\left[\widetilde R \delta(1-A f^*(H))\phi(H)\phi(H)^T\right]=E\left[R \delta(1-A f^*(H))\phi(H)\phi(H)^T\right],$$
but
$$E\left[\widetilde R^2 I(1-A f^*(H)>0)\phi(H)\phi(H)^T\right]
\le E\left[R^2 I(1-A f^*(H)>0)\phi(H)\phi(H)^T\right],$$
using the new weights $\widetilde R_i$ should yield a more accurate estimated and consistent rule in a large sample sense. As a remark, the above conclusion is independent of the choice of large margin loss functions. In other words, if we replace the hinge loss $(1-Af(H))^+$ by other types of large margin losses, using the new weights $\widetilde R_i$ still leads to a more efficient estimation of the optimal treatment rule.

Note that O-learning in (\ref{iowlnew}) is invariant to rescaling $\widetilde R_i$ by any positive constant. Thus, we can always assume $|\widetilde R_i|\le 1$ after a proper scaling. We further observe from (\ref{optima}) that the derived rule remains consistent if we replace $\widetilde R_i$ by $\widetilde R_i-s(H_i)$ for any measurable function $s(H)$. Particularly, if we choose a good candidate $s(H)$ so that the upper bound of $|R-s(H)|$ is even smaller than $|R|$ conditional on $H$ with a large probability, then the {value loss} due to finite sample estimation may be reduced. One heuristic justification for the latter is based on the proof of Theorem 3.4 in Zhao et al.~(2012), also seen in Steinwart \& Scoverl (2007), where it shows that the upper bound for the value loss due to the derived rule is a monotone function of a constant $C_L$ which is the Lipschitz constant of the weighted loss function, $w_i (1-A_if(H_i))^+$ with respect to $f$ in $L_2(P)$-norm. Thus, if $E[(\widetilde R_i-s(H_i))^2|H_i]$ is less than $E[\widetilde R_i^2|H_i]$, this bound under the weight $w_i=(\widetilde R_i-s(H_i))$ is tighter than the one under the weight $w_i=\widetilde R_i$. In particular, we can choose $s(H_i)$ to be an approximation to $E[\widetilde R_i|H_i]$ and then use $\widetilde R_i-s(H_i)$ as the new weights in (\ref{iowlnew}). In other words,
we propose to minimize
\begin{equation}
\min_{f}n^{-1}\sum_{i=1}^n(1-A_if(H_i))^+\frac{\widetilde R_i-s(H_i)}{\pi_i}+\lambda_n\|f\|^2.
\label{iowlnew2}
\end{equation}
Our subsequent simulation studies show that this modification can sometimes significantly improve the original O-learning.

Finally, by using the new weights $\widetilde R_i-s(H_i)$, the weights in (\ref{iowlnew2}) will have negative values which renders the objective function in (\ref{iowlnew2}) to be non-convex. To fix this issue, since asymptotically the goal of  optimization is to derive a rule minimizing
$$E\left[\frac{\widetilde R-s(H)}{\pi(A,H)}I(A\neq \textrm{sign}(f(H)))\right],$$
which is equivalent to
$$E\left[\frac{|\widetilde R-s(H)|}{\pi(A,H)}I(A\textrm{sign}(\widetilde R-s(H))\neq \textrm{sign}(f(H)))\right]-E\left(\frac{(\widetilde R-s(H))^-}{\pi(A,H)}\right),$$
this suggests that we can minimize
\begin{equation}
\min_{f}n^{-1}\sum_{i=1}^n(1-A_i\textrm{sign}(\widetilde R_i-s(H_i))f(H_i))^+\frac{|\widetilde R_i-s(H_i)|}{\pi_i}+\lambda_n\|f\|^2,
\label{iowlnew3}
\end{equation}
which is a convex optimization problem.
Intuitively, for a subject with large positive outcome $\widetilde R_i-s(H_i)$, the desirable decision rule is to choose the currently observed treatment $A_i$ since $f(H_i)$ is encouraged to have the same sign as $A_i$. In contrast, for a subject with negative outcome $\widetilde R_i-s(H_i)$, the currently observed  $A_i's$ might not be the best treatment choice, so a desirable decision rule  is encouraged to be the alternative treatment $-A_i$.

In summary, we propose the following steps to improve the single stage O-learning:
\\
{\it Step 1}. Identify new weights $\widetilde R_i$ which have the same conditional mean as $R_i$ but smaller variance, and  estimate them consistently as
$\widehat R_i=\widehat g(R_i,A_i,H_i)$ with some data-dependent function $\widehat g(\cdot)$;
\\
{\it Step 2}. Find a function $\widehat s(H_i)$ to approximate $E[\widetilde R_i|H_i]$, for instance, through a regression model of $\widehat R_i$ on $H_i$, and calculate the new weight $\widehat R_i-\widehat s(H_i)$;
\\
{\it Step 3}. Solve the minimization problem in (\ref{iowlnew3}) to estimate $f(H)$, where $\widetilde R_i-s(H_i)$ is replaced by $\widehat R_i-\widehat s(H_i)$.

Computationally, minimizing (\ref{iowlnew3}) can be carried out using an algorithm similar to support vector machine. Specifically,  it is equivalent to solving a convex optimization with slack variables $\xi_i's$:
\begin{equation}
\begin{split}
&\min_{f,\xi_i}\frac{1}{2}\|f\|^2+{\cal C}\sum_{i=1}^n\xi_i\frac{|\widehat R_i-\widehat s(H_i)|}{\pi_i} \\
\textrm{subject to }& \xi_i\ge0, \textrm{sign} (\widehat R_i-\widehat s(H_i)) A_if(H_i)\ge 1-\xi_i, \forall i=1,...,n.
\label{primal}
\end{split}
\end{equation}
The decision function $f(h)$ could be linear $f(h)=h^T\beta +\beta_0,$
or a linear combination of basis functions of a functional space ${\cal H}$: $f(h)=\phi(h)^T\theta+\theta_0, \textrm{ where } \phi(h)=(\phi_1(h),\dots,\phi_m(h),\dots),$ and $\phi$ can have finite or infinite dimensions. In both cases, 
we solve for parameters $\theta$ and $\theta_0$ via the dual problem:
\begin{equation*}
\max_{\alpha_i's}\sum_{i=1}^n\alpha_i-\frac{1}{2}\sum_{i=1}^n\sum_{j=1}^n\alpha_i\textrm{sign}(\widehat R_i-\widehat s(H_i))A_iK(H_i,H_j)A_j\textrm{sign}(\widehat R_i-\widehat s(H_i))\alpha_j
\end{equation*}
{subject to } $ 0\le\alpha_i\le {\cal C}{|\widehat R_i-\widehat s(H_i)|}/{\pi_i},$
where $K(\cdot,\cdot)$ is a kernel function. The coefficient $\theta$ is computed as $$\widehat{\theta}=\sum_{i=1}^n\alpha_i\textrm{sign}(\widehat R_i-\widehat s(H_i))A_i\phi(H_i)$$
and the intercept parameter $\theta_0$ can be computed from the Karush-Kuhn-Tucker (KKT) conditions. Finally, kernel tricks can be used to estimate nonlinear function $f(h)$ when $K(\cdot,\cdot)$ is chosen to be some nonlinear kernel such as the Gaussian kernel.

\subsection{Simple Augmented Multistage O-Learning (AMOL)}

Here we apply the general idea of improving  O-learning to multiple-stage studies. Since DTRs aim to maximize the expected cumulative rewards across all stages, the optimal treatment decision at the current stage must depend on subsequent decision rules and future clinical outcomes or rewards under those rules. This observation motivates us to use a backward recursive procedure similar to the backward induction in Q-learning and O-learning in \citeA{zhao2014}.
At each stage, we will follow Steps 1-3 as given in the previous section with carefully chosen weights. Specifically, the new weights in Step 1 is based on imputing the outcomes for subjects who do not follow the estimated optimal rule in latter stages through a regression model so as to use information from all the subjects at all stages. The imputation is realized in a manner similar to the augmented inverse probability weighted complete-case (AIPWCC) method in the missing data literature \cite{tsiatis2006}, where we robustly estimate the optimal treatment regimes even if the regression imputation model is misspecified.

The proposed augmented multistage outcome-weighted learning (AMOL) is described as follows. Start from the last stage $K$. Following the proposed improved single-stage O-learning, we first fit a parametric regression model to estimate $E[R_{iK}|H_{iK}]$ as $\widehat s_K(H_{iK})$, and then solve $$\widehat{f}_K=\textrm{argmin}_{f_K}n^{-1}\sum_{i=1}^n\frac{(1-\textrm{sign}(R_{iK}-\widehat s_K(H_{iK}))A_{iK}f_K(H_{iK}))^+|R_{iK}-\widehat s_K(H_{iK})|}{\pi_{iK}(A_{iK},H_{iK})}+\lambda_{nK}\|f_K\|^2.  $$
Denote the optimal decision at the last stage as $\widehat {\cal D}_K(h)=\textrm{sign}(\widehat{f}_K(h))$.

To identify the optimal treatment rule for stage $K-1$, we need to determine the potential future reward had the subject followed the optimal treatment at stage $K$. Clearly, for subject $i$ whose assigned treatment $A_{iK}$ is the same as the optimal treatment, $\widehat{\cal D}_K(H_{iK})$, the optimal reward value is the observed $R_{iK}$ since the observed treatment assignment at the last stage is optimal. For subjects who did not receive the optimal treatment $\widehat{\cal D}_K(H_{iK})$, the observed $R_{iK}$ is no longer the maximal reward following the optimal treatment rule. However, using all subjects and their health history information at this stage, one can fit a regression model for $R_{iK}$ given $(A_{iK}, H_{iK})$. This model can be either parametric or fully nonparametric. From this estimated regression model,  the optimal reward expected at the last stage given history information $H_K=h$ can be obtained if the regression model is correctly specified. That is, we can obtain the expected maximal reward at stage $K$ as
\begin{equation}
\widehat{g}_K(H_K)=\max_{a\in \{-1,1\}}\widehat E[R_K|A_K=a, H_K=h],\label{reg}
\end{equation}
where $\widehat E[\cdot]$ denotes the predicted mean from fitting a regression model of $R_k$ given $A_k$ and $H_k$ (including their interactions).

The estimated optimal reward $\widehat{g}_K(H_K)$ for a subject  in (\ref{reg})  is only valid under the assumption that the regression model used to estimate the conditional mean in (\ref{reg}) is correctly specified. Therefore, in order to simultaneously use all the subjects and be protected from  misspecification of the regression model, we construct pseudo-outcomes for doubly robust estimation using a similar strategy as in the missing data literature. Considering subjects who did not follow the optimal treatment assignment at stage $K$ as missing outcomes, we form pseudo-outcomes
\begin{equation}
\widehat Q_{iK}=\frac{I(A_{iK}=\widehat{\cal D}_K(H_{iK}))}{\pi_{K}(A_{iK},H_{iK})} R_{iK}
-\frac{I(A_{iK}=\widehat{\cal D}_K(H_{iK}))-\pi_{K}(A_{iK},H_{iK})}{\pi_{K}(A_{iK},H_{iK})} \widehat{g}_K(H_K).\label{qik}
\end{equation}
The first term in $\widehat Q_{iK}$ is the inverse probability weighted complete-case estimator since only when $I(A_{iK}= \widehat{\cal D}_K(H_{iK}))=1$ would the observed outcome $R_{iK}$ be deemed as non-missing. The second term is an augmentation term incorporating  contributions from the subjects who did not receive the optimal treatment assignments. For these subjects,  their pseudo-outcomes reduce to the conditional expectations $\widehat{g}_K(H_K)$.
We take $R_{i,K-1}+ \widehat Q_{iK}$ as the outcome  $\widetilde R_{ik}$ in a single-stage O-learning and follow Steps 2-3 in the improved O-learning. Denoting the weights for stage $K-1$ as $$\widehat{W}_{i,K-1}=\frac{R_{i,K-1}+ \widehat Q_{iK}-\widehat s_{K-1}(H_{i,K-1})}{\pi_{i,K-1}(A_{i,K-1},H_{i,K-1})},$$ where $\widehat s_{K-1}(H_{i,K-1})$ is the estimated mean of $R_{i,K-1}+\widehat Q_{iK}$ given $H_{i,K-1}$ based on a  regression model, we can estimate the optimal rule for stage $K-1$ by minimizing
\begin{equation}
n^{-1}\sum_{i=1}^n(1-\textrm{sign}(\widehat{W}_{i,K-1})A_{i,K-1}f_{K-1}(H_{i,K-1}))^+|\widehat{W}_{i,K-1}|
+\lambda_{n,K-1}\|f_{K-1}\|^2. \end{equation}
The optimal rule at stage $K-1$ is then  $\widehat{\cal D}_{K-1}(h)=\textrm{sign}(\widehat{f}_{K-1}(h))$.

The above procedure can be continued to other stages. Specifically, at the $(k-1)^\textrm{th}$-stage, we first regress $R_k+\widehat{g}_{k+1}(H_{k+1})$ on $A_k$, $H_k$ and their interactions to compute
$$\widehat{g}_{k}(h)=\max_{a\in \{-1,1\}}\widehat E[R_k+\widehat{g}_{k+1}(H_{k+1})|A_{k}=a, H_k=h],$$
where $\widehat E[\cdot]$ is the predicted mean from the regression model of $R_k+\widehat{g}_{k+1}(H_{k+1})$ on ($A_k, H_k, A_kH_k)$.
Second,  we compute the pseudo-outcomes
$$\hspace{-2in}\widehat Q_{ik}=\frac{\prod_{j=k}^KI(A_{ij}=\widehat{\cal D}_j(H_{ij})(\sum_{j=k}^KR_{ij})}{\prod_{j=k}^K\pi_j(A_{ij}, H_{ij})} $$$$-\frac{\prod_{j=k}^KI(A_{ij}=\widehat{\cal D}_{j}(H_{ij})) -\prod_{j\ge k}\pi_j(A_{ij}, H_{ij})}
{\prod_{j\ge k}\pi_j(A_{ij}, H_{ij})}\widehat{g}_{k}(H_{i,k}).$$
Following Steps 2-3 in the improved O-learning, we denote the weight for stage $k-1$ as $\widehat{W}_{i,k-1}$, where  $$ \widehat{W}_{i,k-1}=\frac{R_{i,k-1}+ \widehat Q_{ik}-\widehat s_{k-1}(H_{i,k-1})}{\pi_{k-1}(A_{i,k-1},H_{i,k-1})}$$ with $\widehat s_{k-1}(H_{i,k-1})$ being the estimated mean of $(R_{i,k-1}+\widehat Q_{ik})$ given $H_{i,k-1}$. Then we estimate the optimal rule for stage $k-1$ by computing the minimizer,  $\widehat{f}_{k-1}$, of the following function:
\begin{equation}
n^{-1}\sum_{i=1}^n\left(1-\textrm{sign}(\widehat{W}_{i,k-1})A_{i,k-1}f_{k-1}(H_{i,k-1})\right)^+\left|\widehat W_{i,k-1}\right|
+\lambda_{n,k-1}\|f_{k-1}\|^2. \end{equation}
The corresponding optimal rule at stage $k-1$ is  $\widehat{\cal D}_{k-1}(h)=\textrm{sign}(\widehat{f}_{k-1}(h))$. We repeat this procedure for all $K$ stages.

When the regression imputation model is correct, we expect AMOL to improve estimation efficiency by borrowing information from the model-based Q-learning. However, when the imputation model is misspecified, AMOL continues to provide consistent decision functions due to the double robustness (see Theorem 3.3), while Q-learning does not maintain this consistency. This is due to using the augmented inverse probability weighted complete-case estimator. Therefore as long as either the $\pi_{ik}$ or the conditional expected outcome $\widehat{g}_{i,k}$ is estimated consistently, $R_{i,k-1}+Q_{ik}$ is a consistent estimator of the reward under the optimal regimen in future stages. For SMART studies, the randomization probabilities are known by design. Thus, $\pi_{ik}'s$ are always correctly specified, and the double robustness ensures consistency of AMOL.

\subsection{Improve Efficiency of AMOL}

The augmentation (\ref{qik}) in the previous section, although easy to be implemented, may not be the best augmentation for all stages in the context of missing data. {This is because at stage $k$, it does not account for the heterogeneity in the observed treatment patterns  (i.e., ``missing data patterns", or patterns of whether an optimal treatment is received at a future stage) among all subjects who do not take optimal treatments in any future stage. In other words, subjects who ``missed" one stage of optimal treatment in the future stage is weighted the same as subjects who ``missed" more than one stage of optimal treatment.}  Specifically, for any subject $i$ with observed history information $H_{ik}$, the simple AMOL augmentation  estimates the value increment since stage $k$ under the optimal treatment rules, denoted by $Q_{ik}=R_{ik}^{({\cal D}*)}+...+R_{iK}^{({\cal D}*)}$, where $V_{ik}^{({\cal D}*)}$ is the potential value for any variable $V_{ik}$ if the treatment rules are optimal from stage $k$  onwards, i.e., $A_{ij}={\cal D}_{j}^*(H_{ij})$, for all $j= k, k+1, ..., K$, only using information up to stage $k$ (i.e., $H_{ik}$) regardless of the heterogeneity in future stages after stage $k$ (some stages optimal and some stages non-optimal).

We propose a more efficient AMOL taking into account the additional information in observed treatment patterns since stage $k$. Specifically, for $j\ge k$,  let $M_{i,k-1}=0$ and let $M_{ij}=I(A_{ik}={\cal D}_{k}^*(H_{ik}),...,A_{ij}={\cal D}_{j}^*(H_{ij}))$ denote whether subject $i$ follows the optimal treatment regimes from stage $k$ to $j$. Let $C_{ij}=M_{i,j-1}-M_{ij}$ indicate that subject $i$ follows the optimal treatment up to stage $j-1$ but does not in stage $j$.
Corresponding to the optimal rule $({\cal D}_{k}^*,..., {\cal D}_{K}^*)$, the complete data for subject $i$ would be
$(H_{ik},  H_{iK}^{({\cal D}*)}, R_{iK}^{({\cal D}*)})$, where $ H_{iK}^{({\cal D}*)}$  includes all the  potential interim outcomes and health history information up to the last stage $K$ if the optimal treatment rule $({\cal D}_k^*,...,{\cal D}_K^*)$ were implemented from $k$ to $K$. The observed data follow a monotone missing pattern: for subject $i$ with $C_{ij}=1$, we only observe $(H_{ik}, H_{ij}^{({\cal D}*)})$ with $H_{ij}^{({\cal D}*)}$ being a subset of $H_{iK}^{({\cal D}*)}$ (since for  this subject only treatments from stage $k$ to $j$ are optimal); and only subjects with $M_{iK}=1$ have observed complete data. From the theory of Robins (1994), also seen in Tsiatis (2006) and Zhang et al.~(2013),  the optimal estimation of $Q_{ik}$ under the monotone missingness and coarsening at random assumption is to use the following augmentation:
$$ Q_{ik}
=\frac{M_{iK}(R_{ik}+...+R_{iK})}{P(M_{iK}=1|H_{iK})}+\sum_{j=k}^K
\frac{C_{ij}-P(C_{ij}=1|M_{i,j-1}=1, H_{ij})M_{i,j-1}}{P(M_{ij}=1|H_{ij})}  m_{kj}(H_{ij}),$$
where $ m_{kj}(H_{ij})=E[Q_{ik}|H_{ij}^{({\cal D}*)}=H_{ij}]$.
From the sequential randomization, we know
$$P(M_{ij}=1|H_{ij})=\prod_{s=k}^j \pi_s({\cal D}_s^*(H_{is}), H_{is})$$ and
$$P(C_{ij}=1|M_{i,j-1}=1, H_{ij})=1-\pi_j({\cal D}_j^*(H_{ij}), H_{ij}).$$
Finally, from the sequential randomization and sequential consistency assumption for potential outcomes, we obtain
\begin{eqnarray*}
m_{kj}(H_{ij})&=&E[Q_{ik}|H_{ij}^{({\cal D}*)}=h]\\
&=&E[R_{ik}^{({\cal D}*)}+...+R_{i,j-1}^{({\cal D}*)}+Q_{ij}|H_{ij}^{({\cal D}*)}=h, A_{ik}={\cal D}_k^*(H_{ik}),...
A_{i,j-1}={\cal D}^*_{j-1}(H_{i,j-1})]\\
&=&E[R_{ik}+...+R_{i,j-1}+Q_{ij}|H_{ij}=h, A_{ik}={\cal D}_k^*(H_{ik}),...
A_{i,j-1}={\cal D}^*_{j-1}(H_{i,j-1})]\\
&=&R_{ik}+...+R_{i,j-1}+E[Q_{ij}|H_{ij}=h, A_{ik}={\cal D}_k^*(H_{ik}),...
A_{i,j-1}={\cal D}^*_{j-1}(H_{i,j-1})],
\end{eqnarray*}
which can be estimated using $\widehat g_j(h)+R_{i,j-1}+...+R_{ik}$ for subjects with $M_{i,j-1}=1$ as defined in Section 2.2.

To implement the more efficient AMOL, at stage $k-1$, assuming that we have already obtained the optimal rules after this stage, denoted by $\widehat {\cal D}_{k},..., \widehat{\cal D}_K$, define
$$\widehat M_{ij}=I(A_{ik}=\widehat{\cal D}_{k}(H_{ik}), ..., A_{ij}= \widehat {\cal D}_{j}(H_{ij})),$$ and define $\widehat C_{ij}$ similarly. We estimate $P(M_{ij}=1|H_{ij})$  by
$\widehat P(M_{ij}=1|H_{ij})=\prod_{s=k}^j \pi_s(\widehat{\cal D}_s(H_{is}), H_{is})$ and estimate
$P(C_{ij}=1|M_{i,j-1}=1, H_{ij})$ by $\widehat P(C_{ij}=1|M_{i,j-1}=1, H_{ij}) =1-\pi_j(\widehat{\cal D}_j(H_{ij}), H_{ij}).$
Then the pseudo-outcome for $Q_{ik}$ is calculated as
\be
\widehat Q_{ik}
=\frac{\widehat M_{iK}(R_{ik}+...+R_{iK})}{\widehat P(M_{iK}=1|H_{iK})}+\sum_{j=k}^K
\frac{\widehat C_{ij}-\widehat P(C_{ij}=1|M_{i,j-1}=1, H_{ij})\widehat M_{i,j-1}}{\widehat P(M_{ij}=1|H_{ij})} \nonumber\\
\times
\left\{\widehat g_j(H_{ij})+R_{i,j-1}+...+R_{ik}\right\}.\label{general}
\ee
Once $\widehat Q_{ik}$ is obtained, the algorithm to estimate more efficient AMOL proceeds exactly the same as Section 2.3.

As a remark, no matter how $m_{kj}(H_{ij})$ is estimated, the double robustness of the above augmentation ensures the asymptotic unbiasedness of $\widehat Q_{ik}$. Particularly, if we estimate $m_{kj}(H_{ij})$ using $\widehat g_k(H_{ik})$, i.e., discard partially available information for each treatment pattern (or missingness pattern), then after some algebra, we see that the general AMOL algorithm presented here reduces to the simple AMOL in Section 2.3. In other words, the simple AMOL is a particular augmentation to estimate $Q_{ik}$ at stage $k$. It is equivalent to the more efficient AMOL when there is no intermediate information collected, i.e., $H_{ij}=H_{ik}$ (note that the treatment assignment is determined by $H_{ik}$ under the optimal rule).

\section{Theoretical Results}

We first establish the theoretical results for single-stage AMOL ($K=1$) where the weights are obtained following Steps 1-3 in Section 2.3.
Our first theorem provides the risk bound for ${\cal V}(\widehat f)$, where $\widehat f$ minimizes (\ref{iowlnew3}) with
$\widetilde R=\widehat g(R,A,H)$ for a data-dependent function $\widehat g(\cdot)$ and $s(H)=\widehat s(H)$ for some estimated function $\widehat s(\cdot)$.
We assume that $$
\eta^*(x)=\frac{E[R|A=1,X=x]-E[R|A=-1, X=x]}{E[R|A=1,X=x]/\pi(1,x)+E[R|A=-1, X=x]/\pi(-1,x)}+\frac{1}{2}$$ satisfies the following geometric noise exponent condition (c.f., Steinwart and Scovel, 2007):

\begin{definition}[Geometric noise exponent condition ($q, p$)] For a given function $\eta(x)$, we define
 $$\mathcal{O}^+=\left\{x\in \textrm{Supp}(X): 2\eta(x)-1>0\right\}, \ \
 \mathcal{O}^-=\left\{x\in \textrm{Supp}(X): 2\eta(x)-1<0\right\}.$$
 Let $\Delta(x)=d(x, \mathcal{O}^+)$ if $x\in \mathcal{O}^-$, $\Delta(x)=d(x, \mathcal{O}^-)$ if $x\in \mathcal{O}^+$,
 where $d(x,{\cal O})$ is the distance of $x$ from a set ${\cal O}$.
Then $\eta(x)\in (0,1)$ is assumed to satisfy condition GNE($q$) if
 there exists a constant $C>0$ such that for any $\theta>0,$
$$
E\left[\exp(-\frac{\Delta(X)^2}{\theta})|2\eta(X)-1|\right]\le C\theta^{qp/2},
$$
where $p$ is the dimension of $X$.
\end{definition}
As explained in Zhao et al.~(2012), this geometric noise exponent condition describes the distribution behavior in a neighborhood of the optimal treatment rule. In the situation that optimal treatment assignment is distinct, i.e., $|2\eta^*(x)-1|>
\delta$ for some positive constant $\delta$, $q$ can be any large number.

\medskip

\begin{thm}
\label{thm1} Suppose that the RKHS is the Gaussian kernel space with bandwidth $\sigma_n$. Under the GNE($q$) condition, assume that $\sigma_n=\lambda_n^{-1/(q+1)p}$,  $\lambda_n\rightarrow 0$ and $n\lambda_n\rightarrow\infty$. Additionally, we assume that there exist some function
$g^*(R,A,H)$ and $s^*(H)$ with finite second moments such that $E[g^*(R,A,H)|A,H]=E[R|A,H]$,
$$P\left\{E[|\widehat g(R,A,H)-g^*(R,A,H)|+|\widehat s(H)-s^*(H)|]\le  c'\tau/n^{\beta}\right\}>1-e^{-\tau}$$
and  $P\left\{\Vert \widehat g\Vert_{\infty}+\Vert \widehat s\Vert_{\infty}
\le  c'\tau/\sqrt{\lambda_{n}}\right\}>1-e^{-\tau}$ for some constants $c'$ and $\beta>0$ .
Then for any $\delta>0$, $0<v\le 2$, there exists a constant $c$ such that for all $\tau\ge 1$,
$$P\left\{{\cal V}(\widehat f)\ge {\cal V}(f^*)-\epsilon_n(\tau)\right\}\ge 1-e^{-\tau},$$
where
$$\epsilon_n(\tau)=c\left[n^{-\beta}+\lambda_n^{-\frac{2}{2+v}+\frac{(2-v)(1+\delta)}{(2+v)(1+q)}}n^{-\frac{2}{2+v}}+\frac{\tau}{n\lambda_n}+\lambda_n^{\frac{q}{q+1}}\right].$$
\end{thm}

This theorem implies that ${\cal V}(\widehat f)\rightarrow {\cal V}(f^*)$ in probability and, moreover,
$${\cal V}(\widehat f)-{\cal V}(f^*)=O_p\left(n^{-\beta}+\lambda_n^{-\frac{2}{2+v}+\frac{(2-v)(1+\delta)}{(2+v)(1+q)}}n^{-\frac{2}{2+v}}+\frac{1}{n\lambda_n}+\lambda_n^{\frac{q}{q+1}}\right).$$
The proof of Theorem 3.2 follows from a similar argument to \cite{zhao2012}.

Next, we establish the properties for the simple AMOL in Section 2.3.
Our first result is to justify the use of augmentation in AMOL by the following theorem whose proof is given in the appendix.

\begin{thm}[]
\label{thm0}

For any function $\mu(h)$ which maps from the history information $H_k$ to the outcome domain,
$$
R_{k-1}+\frac{\prod_{j=k}^KI(A_{j}={\cal D}^*_j(H_{j})(\sum_{j=k}^KR_{j})}{\prod_{j=k}^K\pi_j(A_{j}, H_{j})} -\frac{\prod_{j=k}^KI(A_{j}={\cal D}^*_{j}(H_{j})) -\prod_{j\ge k}\pi_j(A_{j}, H_{j})}
{\prod_{j\ge k}\pi_j(A_{j}, H_{j})}\mu(H_k)$$
is always unbiased for $E[R_{k-1}+Q_k|H_k]$, and its conditional variance given $H_k$ is minimized if and only if
$$\mu(H_k)=E[Q_k|H_k]+\left\{Var(\frac{I(\prod_{j=k}^KI(A_{j}={\cal D}^*_j(H_{j}))}{\prod_{j=k}^K\pi_j(A_{j}, H_{j})}|H_k)\right\}^{-1}$$
$$\times
E\left[\frac{I(\prod_{j=k}^KI(A_{j}={\cal D}^*_j(H_{j}))}{\prod_{j=k}^K\pi_j(A_{j}, H_{j})}\frac{1-\prod_{j=k}^K\pi_j(A_{j}, H_{j})}{\prod_{j=k}^K\pi_j(A_{j}, H_{j})}
(Q_k-E[Q_k|H_k])|H_k\right].$$
Here, $Q_k=\sum_{j\ge k}R_j$ under the optimal decision rule, $A_j={\cal D}_j^*(H_j), j\ge k$. Particularly, when $\pi_j(A_j, H_j)$ is constant for all $j=1,...,K$, the optimal $\mu(H_k)$ is $E[Q_k|H_k]$.
\end{thm}

Theorem 3.3 proves that using augmentation, the simple AMOL enjoys a desirable property in
that it always yields the correct future outcome and value function associated with decision rule ${\cal D}$, even if the regression model from the Q-learning for imputation is misspecified. Thus,  AMOL is expected to yield the correct optimal DTRs even under misspecification. Moreover, if the regression model is correctly specified, then the proposed AMOL can have the smallest variability among all unbiased augmentations. Consequently,  the identified DTR is also least variable.

Our last theorem gives the value loss due to the estimated DTRs in the simple AMOL.
Recall a treatment decision rule for the multi-stage DTR is determined by a sequence of functions $f=(f_1,f_2,..., f_K)$ which maps from $(H_1,...,H_K)$ to $\{-1,1\}^K$.  The value function associated with this treatment decision rule is defined as
$${\cal V}(f_1,...,f_K)=E\left[ \frac{(\sum_{j=1}^K R_j)\Pi_{j=1}^K I(A_j=\textrm{sign}\{f_j(H_j))\}}{\Pi_{j=1}^K\pi_j(H_j,A_j)}\right].$$
Further, define the value function at the $k^\textrm{th}$-stage as
$${\cal V}_k(f_k,...,f_K)=E\left[ \frac{(\sum_{j=k}^K R_j)\Pi_{j=k}^K I(A_j=\textrm{sign}\{f_j(H_j))\}}{\Pi_{j=k}^K\pi_j(H_j,A_j)}\right].$$
Then the following theorem gives the main result regarding ${\cal V}_k(f_k^*,...,f_K^*)-{\cal V}_k(\widehat f_k,...,\widehat f_K)$ for $k=1, ..., K$, where $(f_1^*,...,f_K^*)$ is the theoretical optimal treatment rules as given in \cite{zhao2012}.

\begin{thm}
\label{thm2}
Let $\eta_{k}^*$ be defined the same as $\eta^*$ with the following substitution: $A=A_k, X=H_k$ and substitute $R$ with the summation of $R_k$ and the optimal rewards for all future stages. Assume\\
(C.1) $\eta_k^*$ satisfies the GNE condition with $q=q_k$ and $p=p_k$.\\
(C.2) For $k=1,...,K-1$ and $\tau>0$, there exist some function $g_k^*(H_{k+1})$ and $s_k^*(H_k)$ such that
$$P\left\{E[|\widehat g_k(H_{k+1})-g_k^*(H_{k+1})|+|\widehat s_k(H_k)-s_k^*(H_k)|]\le c'\tau/n^{\beta_k}\right\}>1-e^{-\tau}$$
and $P\left\{\Vert \widehat g_k\Vert_{\infty}+\Vert\widehat s_k\Vert_{\infty}\le c'\tau/\sqrt{\lambda_{n,k+1}}\right\}>1-e^{-\tau}$
for some constants $c'$ and $\beta_k>0$
\\
(C.3) The RKHS is the Gaussian kernel space with bandwidth $\sigma_{nk}=\lambda_{nk}^{-1/(q_k+1)p_k}$ and $\lambda_{nk}$ satisfies $n\lambda_{nk}\rightarrow\infty$.\\
Then it holds that for any $\delta_k>0, 0<v_k\le 2$, there exists a constant $c$ such that for all $\tau\ge 1$,
$$P\left\{{\cal V}_k(\widehat f_k, ..., \widehat f_K)\ge {\cal V}_k(f_k^*,..., f_K^*)-\sum_{j=k}^K c_0^{-(K-j)} \epsilon_{nj}(\tau)\right\}
\ge 1-(K-j+1)e^{-\tau},$$
where
$$\epsilon_{nk}(\tau)=c\left[\lambda_{nk}^{-\frac{2}{2+v_k}+\frac{(2-v_k)(1+\delta_k)}{(2+v_k)(1+q_k)}}n^{-\frac{2}{2+v_k}}
+\frac{\tau}{n\lambda_{nk}}+\frac{\tau}{n^{\beta_k}}
+\lambda_{nk}^{\frac{q_k}{q_k+1}}\right].$$
\end{thm}
The above condition (C.1) is the standard condition in support vector machine theory and this condition describes the data distribution near the optimal separation boundary at each stage. Condition (C.2) is the assumed condition for the prediction model to estimate $g_k$ and $s_k$. This condition holds naturally for $\beta_k<1/2$ if $g_k$ and $s_k$ are from a parametric working model.
This theorem implies that ${\cal V}_k(\widehat f_k, ..., \widehat f_K)\rightarrow {\cal V}_k(f_k^*,..., f_K^*)$ in probability. Furthermore, the convergence rate of this approximation depends on the separability of the optimal treatment regimes, which is reflected by $q_k$ in the geometric noise exponent condition, and the approximation of the weights in terms of $\beta_k$ from all stages since stage $k$.
A proof of this theorem is provided in the Appendix.

Finally, for the general AMOL given in Section 2.4, since the only difference from the simple AMOL is the use of $\widehat Q_{ik}$ for $k=1,...,K$, both Theorems 3.3 and 3.4 hold. In particular, the result of Theorem 3.3 follows from the double robustness property given in Theorem 10.4 in Tsiatis (2008). Furthermore, their result shows that no matter what $m_{kj}(H_j)$ is used in the augmentation, the obtained $Q_{ik}$'s estimate has smaller variability as compared to the inverse probability weighted estimator. That is, the general AMOL should lead to more efficient estimation of the optimal treatment rule than the O-learning.  The proof for the results in Theorem 3.4 remains the same except that the approximation of  $\widehat Q_{ik}$ to $Q_{ik}$ depends on the approximation errors of $\widehat g_{kj}$ to its limit $g_{kj}^*$ for $j\ge k$, which does not affect the risk bound as given in Theorem 3.4.

\section{Simulation Studies}

We conducted extensive simulation studies to compare simple and more efficient AMOL with existing approaches by the value function  (reward) of the estimated treatment rules. Particularly, we compared four methods: (a) Q-learning based on a linear regression model with lasso penalty; (b) O-learning as in  \citeA{zhao2014};
(c) AMOL1, the simple AMOL introduced in section 2.3, where linear regression models with lasso penalty were used to estimate $\widehat s_k$ and $\widehat g_k$;
(d) AMOL2, the more efficient AMOL introduced in section 2.4, where linear regression models with lasso penalty were used to estimate $\widehat s_k$ and $\widehat g_k$.
Additionally, in both Q-learning and the estimation of $\widehat g_k$, the interactions between treatment and health history were included in the regression models.
Our simulation settings imitated either a  two-stage or a four-stage trial.  In the following, we only report the results from the four-stage settings while we provide the results from the two-stage designs in the Supplementary Material.

In the first four-stage setting, we considered $20$ feature variables simulated from a multivariate normal distribution, where the first ten variables $X_1,\cdots, X_{10}$ had a pairwise correlation of $0.2$, the rest $10$ variables were uncorrelated, {and the variance for $X_j$'s is 1}.
The reward functions were generated as follows:
\bse
R_1=X_1A_1+\mathcal{N}(0,1); \quad
R_2=(R_1+X_2^2+X_3^2-0.8)A_2+\mathcal{N}(0,1); \\
R_3=2*(R_2+X_4)A_3+X_5^2+X_6+\mathcal{N}(0,1); \quad
R_4=(R_3-0.5)A_4+\mathcal{N}(0,1).
\ese
The randomization probabilities of treatment assignment at each stage were allowed to depend on the feature variables through
\bse
P(A_1=1|H_1)=\frac{1}{1+\exp(-0.5 X_1)};\ \ P(A_2=1|H_2)=\frac{1}{1+\exp(0.1 R_1)}; \\
P(A_3=1|H_3)=\frac{1}{1+\exp(0.2 X_3)};\ \ P(A_4=1|H_4)=\frac{1}{1+\exp (0.2 X_4)}.
\ese
In the second four-stage setting, we imitated a situation where the whole population consisted of a finite number of latent subgroups for which the optimal treatment rule was the same within the group. Specifically, we assumed that there were 10 latent groups, labeled as 1 to 10. For subject in group $l$, given a randomly assigned treatment sequence $(A_1,A_2,A_3,A_4)$, the rewards were generated from $R_1=R_2=R_3=0$ and $R_4=\sum_{j=1}^4A_j A_{jl}^*+N(0,1)$, where $A_{jl}^*= 2(\left \lfloor\frac{l}{2^(j-1)}\right\rfloor\mod2)-1$.
Therefore, for any subject from group $l$, the optimal treatment rule should be $(A_{1l}^*,...,A_{4l}^*)$. However, in practice, we did not observe the group label; instead, we observed feature variables which might be informative of the latent group. In this simulation study, the feature variables consisted of $(X_1,...,X_{30})$ from a multivariate normal distribution with the same correlation matrix as the previous setting. Furthermore, $X_1,...,X_{10}$ had the same mean value $\mu_l$ which was a group-specific constant generated from $N(0,5)$, but the mean of $X_{11},...,X_{30}$ were all zeros. In other words, only $X_1,...,X_{10}$ were informative of the group label but the rest variables were pure noises. 

For each data, we applied Q-learning, O-learning and our proposed methods (AMOL1 and AMOL2) to estimate the optimal rule. At each stage $k$, the patient's health history $H_k$ contained feature variables up to stage $k$, treatment assignments received  at all previous stages, interaction between  treatment and feature variable at each previous stages, and interim outcomes at previous stages. Additionally,
linear kernel was used for O-learning, AMOL1, and AMOL2, and the cost parameter was selected by four-fold cross validation.
For each method, the value function corresponding to the estimated optimal rule was computed using expression (\ref{valuefunction}) from the empirical average of a large independent test data  with sample size $20,000$.

\begin{figure}
\centering\vskip -.4in
\includegraphics[width=0.8\textwidth,height=0.45\textheight]{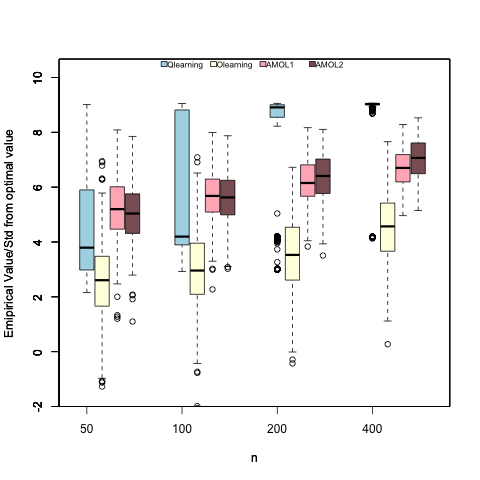}\vskip -.15in
\caption{Simulation setting 1 with four-stage design (optimal value=$10.1$)}
\label{fig1}
\includegraphics[width=0.8\textwidth,height=0.45\textheight]{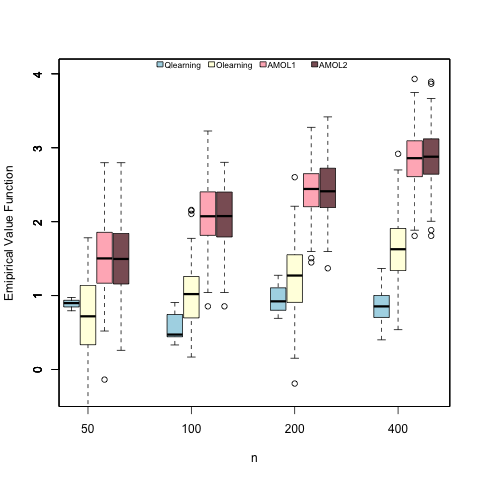}\vskip -.15in
\caption{Simulation setting 2 with four-stage design (optimal value=$4$)}
\label{fig2}
\end{figure}

The results from 500 replicates are presented in Figures \ref{fig1} and \ref{fig2} and Table \ref{simutable}.
In the first setting, we observe that Q-learning incorporating lasso variable selection has high variability of the value function in small sample ($n=$50), but becomes significantly better with increasing sample sizes (e.g., 200 and 400) since the true model for the reward function is based on a regression model. Comparatively, AMOL1 and AMOL2 both have a significant improvement over  O-learning in all sample sizes in terms of larger value and smaller variability due to the data augmentation in the propose methods. They also  outperforms Q-learning in small sample sizes ($n=50$ and $n=100$). Moreover,  when the sample size increases, AMOL2 tends to yield slightly larger mean and median value than AMOL1. This may be because AMOL2 makes use of better augmentation so leads to more accurate estimate of the optimal rule.
In the second setting, since the rewards are generated from a latent mixture model, the regression model in Q-learning is very likely to be misspecified, and thus it performs poorly under all sample sizes (only achieves a median of 0.85 when $n=400$, compared to the actual optimal value  4).
Instead, both AMOL1 and AMOL2 outperform O-learning and Q-learning in all cases, and achieves a value median value of 2.88 for a sample size of 400.  AMOL1 and AMOL2 perform similarly in terms of mean and median values. Similar conclusion is observed in the two-stage simulation studies reported in the Supplementary Material.

\begin{table}
\centering
\caption{Mean and median of empirical value function for 3 simulation scenarios evaluated on independent testing data.}
\footnotesize
\begin{tabular}{cllllllll}
\toprule\toprule
&&&\multicolumn{3}{c}{Simulation setting 1 (optimal value $10.1$)}&&&\\
\midrule
    $n$  &\multicolumn{2}{c}{Q-learning} & \multicolumn{2}{c}{Olearning}& \multicolumn{2}{c}{AMOL1}& \multicolumn{2}{c}{AMOL2}\\
     &Mean (Std) & Median & Mean (Std) &Median &Mean (Std) & Median&Mean (Std) & Median\\\midrule
 50 & 4.672(2.154) & 3.793 & 2.605(1.490) & 2.607 & 5.228(1.156) & 5.197 & 5.030(1.075) & 5.038 \\
 100 & 6.004(2.557) & 4.196 & 3.001(1.453) & 2.959 & 5.704(0.952) & 5.680 & 5.634(0.946) & 5.625 \\
 200 & 7.774(2.135) & 8.908 & 3.521(1.362) & 3.529& 6.242(0.817) & 6.151 & 6.417(0.812) & 6.406 \\
 400 & 8.736(1.117) & 9.032 & 4.521(1.279) & 4.568 & 6.685(0.664) & 6.700 & 7.034(0.706) & 7.066 \\
\midrule\midrule
&&&\multicolumn{3}{c}{Simulation setting 2 (optimal value $4$)}&&&\\
\midrule
    $n$  &\multicolumn{2}{c}{Q-learning} & \multicolumn{2}{c}{Olearning}& \multicolumn{2}{c}{AMOL1}& \multicolumn{2}{c}{AMOL2}\\
     &Mean (Std) & Median & Mean (Std) &Median &Mean (Std) & Median&Mean (Std) & Median
\\\midrule
 50 & 0.890(0.090) & 0.900& 0.726(0.536) & 0.721 & 1.523(0.523) & 1.504 & 1.502(0.510) & 1.496\\
 100 & 0.563(0.217) & 0.473 & 1.027(0.431) & 1.021 & 2.052(0.458) & 2.074& 2.049(0.425) & 2.076 \\
 200 & 0.954(0.241) & 0.923 & 1.251(0.498) & 1.272 & 2.428(0.365) & 2.443 & 2.451(0.388) & 2.411 \\
400 & 0.861(0.226) & 0.855& 1.639(0.446) & 1.628 & 2.846(0.353) & 2.859 & 2.879(0.351) & 2.879 \\
 \bottomrule\bottomrule
\end{tabular}
\label{simutable}
\end{table}

\section{Application}
\begin{figure}

\centering
\includegraphics[angle=0,width=4.5in, totalheight=3in]
{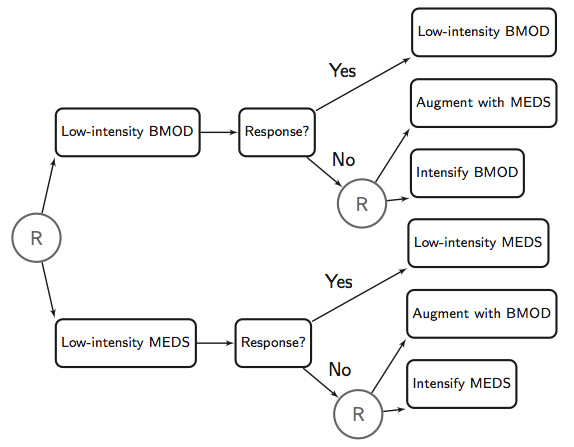}
\caption{Design of Adaptive Pharmacological Behavioral Treatments for Children with ADHD Trial; BMOD, behavior modification; MEDS, low-dose medication.}\label{adhd}
\end{figure}

Our first application is based on a  two-stage SMART for children affected by ADHD  \cite{pelham2008,workshop}.  
Interventions in the ADHD trial were different doses of methamphetamine (MED) and different intensities of behavioral modification (BMOD). As shown in Figure~\ref{adhd}, children were randomly assigned to begin with low-intensity behavioral modification or with low-dose medication. This stage lasted for two months, after which the Impairment Rating Scale (IRS) and the individualized list of target behaviors (ITB) were used to assess each child's response to initial treatment. Children who responded continued to receive the initial low intensity treatment, and children who did not respond were re-randomized to either intensify the initial treatment or switch to the other type of treatment. The primary outcome of the study was a school performance score measured at the end of study which ranges from $1$ to $5$.
For the ADHD data, there were $150$ patients in total, $51$ in-remission after the first stage, and $99$ entered second stage randomization. The randomization probabilities were $0.5$ at both stages.

We compared the performance of Q-learning, O-learning and AMOL when analyzing the ADHD data, nothing that AMOL1 and AMOL2 were equivalent for a  two-stage trial.  The feature variables for the first stage included ODD (Oppositional Defiant Disorder) Diagnosis, baseline ADHD score,  race (white) and prior medication prescription. There were two intermediate variables, months to non-response and adherence to the first stage treatment, to be included as the feature variables for the second stage in addition to those from the first stage.  The final reward was the school performance score at the end of second stage.

For comparison, we randomly partitioned data into a training set and a testing set with sample size ratio 5:1. We then estimated the optimal treatment rule using each method using the training data, and then evaluated the predicted value function in the test data. Particulary, the empirical value function was calculated using expression (\ref{valuefunction}), where $D_k$'s were replaced by the estimated optimal rules. 

LASSO penalization was applied in both Q-learning and the regression steps of AMOL. Here we considered for the remissed patients, continue the current treatment is their optimal treatment. So to learn the optimal rule for second stage all three methods used the nonremissed patients for second stage. Q-learning was not weighted, O-learning and AMOL were weighted by randomization probability $\pi_2=0.5$.  For the first stage, Q-learning used all the patients with no weights. All patients had their final outcome at the end of second stage. O-learning considered the remissed and nonremissed whose observed treatment 2 was the same as estimated optimal. The outcome for the remissed patient was the final outcome weighted by $\pi_1$, the outcome for the non-remissed patients were weighted by $\pi_1*\pi_2$; and AMOL considered the all the patients with complete data for the first stage, the remissed patients were considered in the group whose observed treatment were the same with optimal. AMOL pseudo-outcome was computed and the weights were first stage randomization probability $\pi_1=0.5$

\begin{figure}
\centering
\includegraphics[angle=0,width=3in, totalheight=3in]
{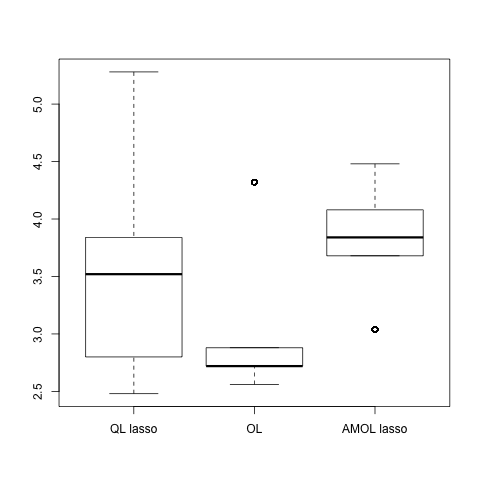}
\caption{Mean and empirical standard error of the value function (school performance score, higher score desirable) based on $100$ repetitions of $2$-fold cross validation using ADHD data}\label{adhd1}
\end{figure}
\begin{figure}
\centering
\includegraphics[angle=0,width=1.5in, totalheight=2.5in]
{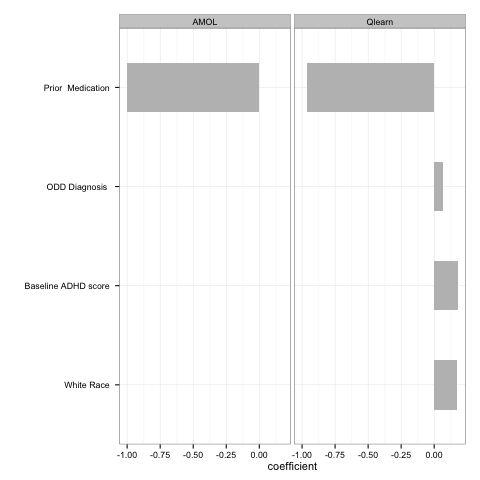}
\includegraphics[angle=0,width=1.5in, totalheight=2.5in]
{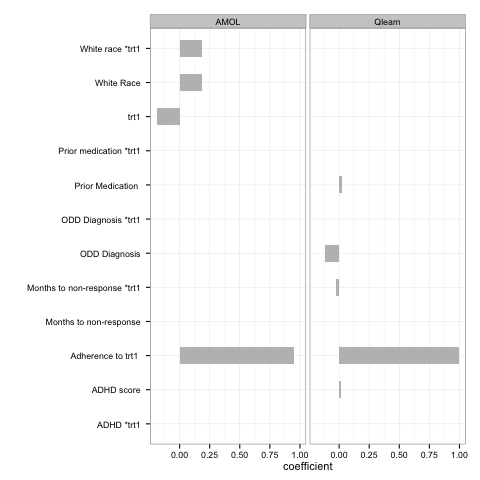}
\caption{Normalized coefficients of the stage 1 tailoring variables (left panel) and stage 2 tailoring variables (right panel) obtained by AMOL and Q-learning}\label{adhd2}
\end{figure}

The box plot of the value estimates from 500 times of randomly splitting training and testing data is presented in Figure \ref{adhd1}. The best one-size-fits-all rules yielded a score of $3.51$ points for the treatment rule beginning with BMOD and then augmenting with MED; and the worst one-size-fits-all rule was to begin with BMOD and intensify BMOD, which yielded a value of $2.65$ points. However, AMOL achieved an average mean value of $3.82$, which is higher than both O-learning (3.04) and Q-learning (3.58).
To visualize the importance of each tailoring variable on the optimal DTR estimated by AMOL and Q-learning, we present the normalized coefficients of the baseline and stage 1 tailoring variables  in Figure \ref{adhd2}. Based on AMOL, the most significant baseline tailoring variable  is  medication prior to enrollment, while all the other variables had negligible magnitudes ($<0.0001$). Q-learning also reveals prior medication as the most important variable but also yields additional small effects for some other variables. Interestingly, for stage 2, both methods identify adherence to stage 1 treatment as the most important tailoring variable. The estimated rule from AMOL suggests that children  who do not have prior exposure to medication before the trial should start with behavioral modification, while those who have prior medication exposure should start with the study  medication.  For stage $2$,  children who adhere to their initial treatment but do not respond should continue with an intensified version of the same treatment, while children who do not adhere  should augment their initial assignment with the alternative treatment.
We also analyze data from STAR*D study \cite{rush2004} and the results are presented in the Online Supplementary Material.

\section{Discussion}

In this work,  we proposed AMOL to estimate DTRs through doubly robust augmentation based on two popular methods: Q-learning and O-learning.  
Q-learning as a regression-based approach identifies optimal DTRs through fitting parametric regression models; while the original O-learning targets the value function directly and introduces nonparametric decision rules through kernel functions. Proposed AMOL draws efficiency from Q-learning to improve O-learning while maintaining consistency. In trials such as EMBARC, it is difficult to identify apriori which variables may serve as tailoring variables for  treatment response.  AMOL may be superior in such settings with large number of non-treatment-differentiating noise variables.  {The appendix in a related working paper in \citeA{laan2014} described the idea of double robust estimation, but without explicit support vector machine implementation or the consideration of convergence rates of the learning algorithm. It would be interesting to further explore how to combine the advantages of AMOL and their super learning  improving the estimation of personalized treatment regimes.}

Clinicians may be interested in ranking the most important variables to predict patient heterogeneity to treatment. Biomarkers that could signal patients' heterogeneous responses to various interventions are especially useful as tailoring variables. This information can be used to design new intervention arms in future confirmatory trials and facilitate discovering new knowledge in medical research. Variable selection may help construct a less noisy rule and avoid over-fitting. Although AMOL leads to a sparse DTR in the ADHD and STAR*D examples, a future research topic is to investigate methods that perform automatic variable selection in the outcome-weighted learning framework.

Our current framework can easily handle nonlinear decision functions by using nonlinear kernels. Such kernels may improve performance for high-dimensional correlated tailoring variables.  Kernel techniques perform automatic dimension reduction through the kernel functions turning a $p\times p$ matrix into an $n\times n$ matrix, where $p$ is the number of feature variables and $n$ is the sample size. 
In addition, it is possible to consider other kinds of decision functions such as decision trees to construct DTRs that are highly interpretable.

In clinical studies, combinations of various outcomes  may be of interest \cite{teixeira2009}. For example, alleviation of symptoms may be considered in conjunction with increased quality of life and functioning, time to response, and reduction of side effect.  Therefore, it may be insufficient to represent all  information in a one-dimensional outcome. It is worth exploring machine learning methods to deal with multi-dimensional responses or multi-dimensional value functions. Lastly, in some clinical studies there may be more than two treatment options at a given stage. AMOL can be extended to handle more than two treatment groups through multi-category classification.

%
%
%


\bigskip
\begin{center}
\bf APPENDIX
\end{center}

\noindent\textbf{Proof of Theorem 3.2}.

We let
$${\cal V}(f;g,s)=E\left[\frac{|g(R,A,H)-s(H)|}{\pi(A,H)}I(A\textrm{sign}(g(R,A,H)-s(H))=\textrm{sign}(f(H)))\right],$$
and
$${\cal R}_{\phi}(f; g, s)=E\left[\frac{|g(R,A,H)-s(H)|}{\pi(A,H)}(1-A\textrm{sign}(g(R,A,H)-s(H))f(H))^+\right].$$
Clearly, if $\widetilde f$ maximizes ${\cal V}(f;g^*,s^*)$, following the same argument in Section 2.2, we know $\widetilde f$ maximizes
$E\left[(g^*(R,A,H)-s^*(H))I(A= \textrm{sign}(f(H)))/\pi(A,H)\right]$ which is equivalent to $$E\left[g^*(R,A,H)I(A= \textrm{sign}(f(H)))/\pi(A,H)\right]-E[s^*(H)].$$ Since $E[g^*(R,A,H)|A,H]=E[R|A,H]$. We conclude that $\widetilde f$ maximizes $$E\left[RI(A\neq \textrm{sign}(f(H))/\pi(A,H)\right],$$ i.e., ${\cal V}(f)$. This implies that $\widetilde f$ gives the same treatment rule as $f^*$.
We define $\check f$ to maximize ${\cal V}(f; \widehat g, \widehat s)$. Consider the event $\left\{E[|\widehat g(R,A,H)-g^*(R,A,H)|+|\widehat s(H)-s^*(H)|]\le c'\tau/n^{\beta}\right\}$ which has probability at least $1-e^{-\tau}/2$. In this event, we obtain
\begin{eqnarray*}
{\cal V}(f^*)-{\cal V}(\widehat f)&=&{\cal V}(f^*;,g^*,s^*)-{\cal V}(\widehat f; g^*, s^*)\\
&\le& {\cal V}(f^*; \widehat g, \widehat s)-{\cal V}(\widehat f; \widehat g, \widehat s)+c\tau n^{-\beta}\\
&\le&{\cal V}(\check f; \widehat g, \widehat s)-{\cal V}(\widehat f; \widehat g, \widehat s)+c\tau n^{-\beta}\\
&\le&
{\cal R}_{\phi}(\widehat f; \widehat g, \widehat s)-{\cal R}_{\phi}(\check f; \widehat g, \widehat s)+c\tau n^{-\beta},
\end{eqnarray*}
where the last step follows the excess risk result from Theorem 3.2 of Zhao et al. (2012)
but the weight is taken to be $|\widehat g(R,A,H)-\widehat s(H)|/\pi(A,H)$.
The remaining proof then follows the same arguments in the proof of Theorem 3.4 in Zhao et al. (2012) except that their loss function is replaced by
${|\widehat g(R,A,H)-\widehat s(H)|}I(A\textrm{sign}(\widehat g(R,A,H)-\widehat s(H))=\textrm{sign}(f(H)))/{\pi(A,H)}.$ Using their result, we obtain with at least probability $1-e^{-\tau}/2$,
$$
{\cal R}_{\phi}(\widehat f; \widehat g, \widehat s)-{\cal R}_{\phi}(\check; \widehat g, \widehat s)
\le c\left[\lambda_n^{-2/(2+\nu)+(2-\nu)(1+\delta)/[(2+\nu)(1+q)]}n^{-2/(2+\nu)}+\tau/(n\lambda_n)+\lambda_n^{q/(q+1)}\right].$$
Combining the above results, we obtain Theorem 3.2.

\noindent\textbf{Proof of Theorem 3.3}.

For abbreviation, we let ${\cal A}_k=\left\{A_j={\cal D}_j^*(H_j), j\ge k\right\}$ and $p({\cal A}_k)=\prod_{j\ge k}\pi_j(A_j,H_j)$.
By sequential conditional expectations, we obtain
$E[\prod_{j=k}^K I({\cal A}_k)/p({\cal A}_k)|H_k]=1$. Furthermore,
$\sum_{j=k}^KR_j=Q_k$ when $A_j={\cal D}_j^*(H_j), j\ge k$.  Thus, the unbiasedness holds.
To show the second part of the theorem, we note that its conditional variance given $H_k$ is
$$Var(\frac{I({\cal A}_k)}{p({\cal A}_k)} Q_k\Big |H_k)-2E\left[\frac{I({\cal A}_k)}{p({\cal A}_k)}\frac{I({\cal A}_k)-p({\cal A}_k)}{p({\cal A}_k)}
(Q_k-E[Q_k|H_k])|H_k\right](\mu(H_k)-E[Q_k|H_k])$$
$$
+Var(\frac{I({\cal A}_k)}{p({\cal A}_k)}|H_k)(\mu(H_k)-E[Q_k|H_k])^2.$$
Thus it is clear that this variance is minimized when we choose
$$\mu(H_k)=E[Q_k|H_k]+\left\{Var(\frac{I({\cal A}_k)}{p({\cal A}_k)}|H_k)\right\}^{-1}
E\left[\frac{I({\cal A}_k)}{p({\cal A}_k)}\frac{1-p({\cal A}_k)}{p({\cal A}_k)}
(Q_k-E[Q_k|H_k])|H_k\right].$$
When $\pi_j(A_j,H_j)$ is constant for any $j$,
$$E\left[\frac{I({\cal A}_k)}{p({\cal A}_k)}\frac{1-p({\cal A}_k)}{p({\cal A}_k)}
(Q_k-E[Q_k|H_k])|H_k\right]=\frac{1-p({\cal A}_k)}{p({\cal A}_k)} E\left[(Q_k-E[Q_k|H_k])|H_k\right]=0.$$
Then the optimal $\mu(H_k)=E[Q_k|H_k]$.

\medskip

\noindent\textbf{Proof of Theorem 3.4}

We define a conditional value function at stage $k$ as
$$U_k(H_{k+1}; f_{k+1},...,f_K)=E\left[\frac{\sum_{j={k+1}}^K R_j\Pi_{j=k+1}^K
I(A_j=\textrm{sign}(f_j(H_j)))}{\Pi_{j=k+1}^K\pi_j(H_j,A_j)}\Big |H_{K+1}\right].$$
Clearly, ${\cal V}_k(f_k,...,f_K)=E\left[I(A_k=\textrm{sign}(f_k(H_k)))U_k(H_{k+1}; f_{k+1},...,f_K)/\pi_k(A_k,H_k)\right].$
In AMOL for the $K$-stage trial, $\widehat f_k$ minimizes
$${\cal R}_{\phi,n}(f; \widehat f_{k+1},...,\widehat f_K)
=n^{-1}\sum_{i=1}^n |\widehat W_{ik}| (1-A_{ik}\textrm{sign}(\widehat W_{ik}) f(H_{ik}))^++\lambda_{nk} \Vert f\Vert^2,$$
where $\widehat W_{iK}=R_{iK}/\pi_K(A_{iK}, H_{iK})$ and for $k<K$,
\begin{eqnarray*}
\widehat W_{ik}&=&\frac{1}{\pi_k(A_{ik},H_{ik})}\left[\frac{I(A_{i,k+1}\widehat f_{k+1}(H_{i,k+1})>0,..., A_{i,K}\widehat f_K(H_{i,K})>0)(\sum_{j=k}^K R_{ij})}{\prod_{j=k+1}^K\pi_j(A_{ij}, H_{ij})}\right.\\
& &\left.
-\frac{I(A_{i,k+1}\widehat f_{k+1}(H_{i,k+1})>0,..., A_{i,K}\widehat f_K(H_{i,K})>0)-\prod_{j=k+1}^K\pi_j(A_{ij}, H_{ij})}{\prod_{j=k+1}^K\pi_j(A_{ij}, H_{ij})}\right.\\
& & \qquad \left.\times \widehat g_k(H_{i,k+1})+R_{ik}-\widehat s_k(H_{ik})\right],
\end{eqnarray*}
where $\widehat g_k(H_{k+1})$ is the estimator for the predicted optimal value function obtained from $Q$-learning.

At the final stage, $\widehat f_K$ is essentially the estimated rule from a single-stage AMOL, so from Theorem 3.2, we have
$$P({\cal V}_K(f_K^*)-{\cal V}_K(\widehat f_K)\le \epsilon_K(\tau))\ge 1-e^{-\tau}, \eqno(A.1)$$
where $\epsilon_K(\tau)=c\left[n^{-\beta_K}+\lambda_n^{-\frac{2}{2+{\cal V}_K}+\frac{(2-{\cal V}_K)(1+\delta_K)}{(2+{\cal V}_K)(1+q_K)}}n^{-\frac{2}{2+{\cal V}_K}}+\tau/(n\lambda_{nK})
+\lambda_{nK}^{\frac{q_K}{q_K+1}}\right]$ for some constant $c$ and the exponent $q_K$ in the GNE condition for the $K$-th stage. Furthermore, we obtain $\Vert \widehat f_K\Vert\le c/\sqrt{\lambda_{nK}}$ for some constant with at least probability $1-e^{-\tau}/4$.

For $k=K-1$, we  note
$${\cal V}_{K-1}(f_{K-1}^*, f_K^*)-{\cal V}_{K-1}(\widehat f_{K-1}, \widehat f_K)$$
$$
\le {\cal V}_{K-1}(f_{K-1}^*, f_K^*)-{\cal V}_{K-1}(f_{K-1}^*, \widehat f_K)+{\cal V}_{K-1}(f_{K-1}^*, \widehat f_K)-{\cal V}_{K-1}(\widehat f_{K-1}, \widehat f_K).$$
On the other hand, since
$$U_{K-1}(H_{K}; f_K^*)\ge U_{K-1}(H_k; \widehat f_K)$$
and
$$E\left[\frac{I(A_{K}f_K^*(H_K)>0)-I(A_K\widehat f_K(H_K)>0)}{\pi_K(A_K,H_K)}|H_K\right]=0,$$
we have
\begin{eqnarray*}
& &{\cal V}_{K-1}(f_{K-1}^*, f_K^*)-{\cal V}_{K-1}(f_{K-1}^*, \widehat f_K)\\
&=&E\left[\frac{I(A_{K-1}=\textrm{sign}(f_{K-1}^*))}{\pi_{K-1}(A_{K-1}, H_{K-1})} \left\{U_{K-1}(H_K; f_K^*)-U_{K-1}(H_K; \widehat f_K)\right\}\right]\\
&\le& c_0^{-1} E\left[ \left\{U_{K-1}(H_K; f_K^*)-U_{K-1}(H_K; \widehat f_K)\right\}\right]\\
&=& c_0^{-1}\left\{{\cal V}_{K}(f_K^*)-{\cal V}_{K}(\widehat f_K)\right\},
\end{eqnarray*}
where $c_0$ is the lower bound for $\pi_{k}(a_k, h_k)$ for $k=1,...,K$.
Hence,
$${\cal V}_{K-1}(f_{K-1}^*, f_K^*)-{\cal V}_{K-1}(\widehat f_{K-1}, \widehat f_K)$$
$$\le c_0^{-1}\left\{{\cal V}_{K}(f_K^*)-{\cal V}_{K}(\widehat f_K)\right\}+{\cal V}_{K-1}(f_{K-1}^*, \widehat f_K)-{\cal V}_{K-1}(\widehat f_{K-1}, \widehat f_K).\eqno(A.2)$$

We define
$$\widetilde f_{K-1}=\textrm{argmax}_f {\cal V}_{K-1}(f, \widehat f_K).$$
Then (A.2) gives
$${\cal V}_{K-1}(f_{K-1}^*, f_K^*)-{\cal V}_{K-1}(\widehat f_{K-1}, \widehat f_K)$$
$$\le c_0^{-1}\left\{{\cal V}_{K}(f_K^*)-{\cal V}_{K}(\widehat f_K)\right\}+\left\{{\cal V}_{K-1}(\widetilde f_{K-1}, \widehat f_K)-{\cal V}_{K-1}(\widehat f_{K-1}, \widehat f_K)\right\}.\eqno(A.3)$$
Note that the second term on the right-hand side of (A.3) is essentially the value loss due to using $\widehat f_{K-1}$ if we restrict to the population where $A_K=\textrm{sign}(\widehat f_K)$.
Thus, if we let
\begin{gather*}
\omega(f,g, s)=\frac{1}{\pi_{K-1}(A_{K-1}, H_{K-1})}
\Big\{I(A_{K}=\textrm{sign}(f(H_{K})))(R_{K})\\
-\frac{I(A_{K}=\textrm{sign}(f(H_{K})))-\pi_{K}(A_{K}, H_{K})}
{\pi_{K}(A_{K}, H_{K})}g(H_{K})+R_{K-1}-s(H_{K-1})\Big\},
\end{gather*}
then for any $f$,
$$E[\omega(\widehat f_K, g_{K-1}^*, s_{K-1}^*)I(A_{K-1}f(H_{K-1})>0)]$$
$$=E\left[\frac{I(A_{K-1}f(H_{K-1})>0,
A_{K}=\textrm{sign}(\widehat f_K(H_{K}))(R_{K-1}+R_{K})}
{\pi_{K-1}(A_{K-1}, H_{K-1})}\right],$$
and thus
\begin{eqnarray*}
0&\le & {\cal V}_{K-1}(\widetilde f_{K-1}, \widehat f_K)-{\cal V}_{K-1}(\widehat f_{K-1}, \widehat f_K)\\
&=&E[\omega(\widehat f_K, g_{K-1}^*, s_{K-1}^*)I(A_{K-1}\widetilde f_{K-1}(H_{K-1})>0)]\\
& &-E[\omega(\widehat f_K, g_{K-1}^*, s_{K-1}^*)I(A_{K-1}\widehat f_{K-1}(H_{K-1})>0)]\\
&=&E[|\omega(\widehat f_K, g_{K-1}^*, s_{K-1}^*)| I(A_{K-1}\textrm{sign}(\omega(\widehat f_K, g_{K-1}^*, s_{K-1}^*))\widehat f_{K-1}(H_{K-1})\le 0)]\\
& &-E[|\omega(\widehat f_K, g_{K-1}^*, s_{K-1}^*)|
I(A_{K-1}\textrm{sign}(\omega(\widehat f_K, g_{K-1}^*, s_{K-1}^*))\widetilde f_{K-1}(H_{K-1})\le 0)].
\end{eqnarray*}
Moreover, using the excess risk result in Theorem 3.2 of Zhao et al. (2012) (also see Barlett et al. 2006), we obtain
$$
{\cal V}_{K-1}(\widetilde f_{K-1}, \widehat f_K)-{\cal V}_{K-1}(\widehat f_{K-1}, \widehat f_K)
\le {\cal R}_{\phi}(\widehat f_{K-1}; \widehat f_K, g_{K-1}^*, s_{K-1}^*)-{\cal R}_{\phi}(\widetilde f_{K-1}; \widehat f_K, g_{K-1}^*, s_{K-1}^*),
\eqno(A.4)$$
where
$${\cal R}_{\phi}(f; f_K, g_{K-1}, s_{K-1})=E[|\omega(f_K, g_{K-1}, s_{K-1})| (1-A_{K-1}\textrm{sign}(\omega(f_K, g_{K-1}, s_{K-1}))f(H_{K-1}))^+].$$

Consider the set $${\cal A}=\left\{E[|\widehat g_{K-1}(H_K)-g_{K-1}^*(H_K)|+|\widehat s_{K-1}(H_{K-1})-s^*(H_{K-1})|]\le c_1\tau/n^{\beta_{K-1}}\right\}.$$ According to (C.2), $P({\cal A})>1-e^{-\tau}/2$. Furthermore, on this set, it is clear that
\begin{eqnarray*}
& &\sup_{f}|{\cal R}_{\phi}(f; \widehat f_K, g_{K-1}^*, s_{K-1}^*)-{\cal R}_{\phi}(f; \widehat f_K, \widehat g_{K-1},
\widehat s_{K-1})|\\
&\le&
c_2E[|\widehat g_{K-1}(H_K)-g_{K-1}^*(H_K)|+|\widehat s_{K-1}(H_{K-1})-s^*(H_{K-1})|]\\
& &+c_2E[|\omega(\widehat f_K, g_{K-1}^*, s_{K-1}^*)| |\textrm{sign}(
\omega(\widehat f_K, \widehat g_{K-1}, \widehat s_{K-1}))-\textrm{sign}(\omega(\widehat f_K, g_{K-1}^*, s_{K-1}^*))|]
\end{eqnarray*}
for some constant $c_2$.
Since if $\textrm{sign}(
\omega(\widehat f_K, \widehat g_{K-1}, \widehat s_{K-1}))-\textrm{sign}(\omega(\widehat f_K, g_{K-1}^*, s_{K-1}^*)\neq 0$ then
$$|\omega(\widehat f_K, g_{K-1}^*, s_{K-1}^*)|\le |\omega(\widehat f_K, g_{K-1}^*, s_{K-1}^*)-\omega(\widehat f_K, \widehat g_{K-1}, \widehat s_{K-1})|$$
$$\le c_2 \left\{|\widehat g_{K-1}(H_K)-g_{K-1}^*(H_K)|+|\widehat s_{K-1}(H_{K-1})-s_{K-1}^*(H_{K-1})|\right\},$$
and we have
$$\sup_{f}|{\cal R}_{\phi}(f; \widehat f_K, g_{K-1}^*, s_{K-1}^*)-{\cal R}_{\phi}(f; \widehat f_K, \widehat g_{K-1}, \widehat s_{K-1})|\le
c_3\tau/n^{\beta_{K-1}}$$
for some constant $c_3$.

Therefore, combining the results from (A.1)--(A.4), we obtain that by re-defining a constant $c$ in $\epsilon_K(\tau)$, with at least $1-3e^{-\tau}/2$ probability,
$$\hspace{-1.5in}{\cal V}_{K-1}(\widehat f_{K-1}, \widehat f_K)\ge {\cal V}_{K-1}(f_{K-1}^*, f_K^*)-c_0^{-1}\epsilon_K(\tau)
-c\frac{\tau}{n^{\beta_{K-1}}}$$
$$-\left\{{\cal R}_{\phi}(\widehat f_{K-1}; \widehat f_K, \widehat g_{K-1}, \widehat s_{K-1})-{\cal R}_{\phi}(\widetilde f_{K-1}; \widehat f_K, \widehat g_{K-1}, \widehat s_{K-1})\right\}.
\eqno(A.5)$$
Hence, to prove the theorem for $K-1$, it suffices to derive a stochastic bound for the last term in the right-hand side of (A.5).

Note that this term is the excess risk if we treat ${\cal R}_{\phi}(f; \widehat f_K, \widehat g_{K-1}, \widehat s_{K-1})$ as the loss function.
Clearly, $\widetilde f_{K-1}$ is the minimizer and based on the definition of $\widehat f_{K-1}$, $\widehat f_{K-1}$ minimizes its corresponding empirical version with regularization:
$${\cal R}_{\phi n}(f; \widehat f_K, \widehat g_{K-1}, \widehat s_{K-1})\equiv n^{-1}\sum_{i=1}^{n}
| \widehat \omega_i|(1-A_{i,K-1} \textrm{sign}(\widehat\omega_i
)f(H_{i,K-1}))^++\lambda_{n,K-1} \Vert f\Vert,$$
where
$$ \widehat \omega_{i}=\frac{1}{\pi_{K-1}(A_{i,K-1}, H_{i,K-1})}
\left\{(R_{i,K})I(A_{iK}=\textrm{sign}(\widehat f_K(H_{iK})))
\right.$$
$$\left.-\frac{I(A_{iK}=\textrm{sign}(\widehat f_K(H_{iK})))-\pi_{K}(A_{iK}, H_{iK})}
{\pi_{K}(A_{iK}, H_{iK})}\widehat g_{K-1}(H_{iK})+R_{i,K-1}-\widehat s_{K-1}(H_{i,K-1})\right\}.$$
The derivation of the stochastic bound follows the same arguments used in the proof of Theorem 3.4 in Zhao et al. (2012). The only difference is that the loss function corresponding to ${\cal R}_{\phi}$, i.e.,
$$l(f)\equiv |\omega(\widehat f_K, \widehat g_{K-1}, \widehat s_{K-1})|(1-
 A_{K-1}\textrm{sign}(\omega(\widehat f_K, \widehat g_{K-1}, \widehat s_{K-1}))f(H_{K-1}))^+,$$
 depends on $\widehat f_K$ and $\widehat g_{K-1}$ as well as $\widehat s_{K-1}$.
Since the latter are all bounded by $c\tau/\sqrt{\lambda_{n,K}}$ with probability at least $1-e^{-\tau}/8$ for a large $c$, the $\epsilon$-entropy number for $l(f)$ is bounded by $c\epsilon^{-v}\sigma_n^{(1-v/2)(1+\delta)}$ for any $v\in (0,2)$ and $\delta>0$. Therefore, from Theorem 5.6 in Steinwart and Scovel (2007) and using conditions (C.1) and (C.3), we conclude that with at least probability $1-e^{-\tau}/2$,
$${\cal R}_{\phi}(\widehat f_{K-1}; \widehat f_K, \widehat g_{K-1}, \widehat s_{K-1})-{\cal R}_{\phi}(\widetilde f_{K-1}; \widehat f_K, \widehat g_{K-1}, \widehat s_{K-1})$$
$$\ge
c\left[\lambda_{n,K-1}^{-\frac{2}{2+{\cal V}_{K-1}}+\frac{(2-{\cal V}_{K-1})(1+\delta_{K-1})}{(2+{\cal V}_{K-1})(1+q_{K-1})}}n^{-\frac{2}{2+{\cal V}_{K-1}}}
+\frac{\tau}{n\lambda_{n,{K-1}}}
+\lambda_{n, K-1}^{\frac{q_{K-1}}{q_{K-1}+1}}\right]. \eqno(A.6)$$

Finally, combining (A.5) and (A.6) yields
$$
P\left\{{\cal V}_{K-1}(f_{K-1}^*, f_K^*)- {\cal V}_{K-1}(\widehat f_{K-1}, \widehat f_K)\le c_0^{-1}\epsilon_{n,K}(\tau)+\epsilon_{n,K-1}(\tau)\right\}\ge 1-2e^{\tau}.$$
By repeating the above proofs from stage $K-2$ through stage 1, we obtain the result in Theorem 3.4.

\bibliographystyle{apacite}
\bibliography{oral}

\begin{thebibliography}{}

\bibitem [\protect \citeauthoryear {%
Almirall%
, Ten~Have%
\BCBL {}\ \BBA {} Murphy%
}{%
Almirall%
\ \protect \BOthers {.}}{%
{\protect \APACyear {2010}}%
}]{%
almirall2010}
\APACinsertmetastar {%
almirall2010}%
\begin{APACrefauthors}%
Almirall, D.%
, Ten~Have, T.%
\BCBL {}\ \BBA {} Murphy, S\BPBI A.%
\end{APACrefauthors}%
\unskip\
\newblock
\APACrefYearMonthDay{2010}{}{}.
\newblock
{\BBOQ}\APACrefatitle {Structural Nested Mean Models for Assessing Time-Varying
  Effect Moderation} {Structural nested mean models for assessing time-varying
  effect moderation}.{\BBCQ}
\newblock
\APACjournalVolNumPages{Biometrics}{66}{1}{131--139}.
\PrintBackRefs{\CurrentBib}

\bibitem [\protect \citeauthoryear {%
Arjas%
\ \BBA {} Andreev%
}{%
Arjas%
\ \BBA {} Andreev%
}{%
{\protect \APACyear {2000}}%
}]{%
arjas2000}
\APACinsertmetastar {%
arjas2000}%
\begin{APACrefauthors}%
Arjas, E.%
\BCBT {}\ \BBA {} Andreev, A.%
\end{APACrefauthors}%
\unskip\
\newblock
\APACrefYearMonthDay{2000}{}{}.
\newblock
{\BBOQ}\APACrefatitle {Predictive inference, causal reasoning, and model
  assessment in nonparametric Bayesian analysis: a case study} {Predictive
  inference, causal reasoning, and model assessment in nonparametric bayesian
  analysis: a case study}.{\BBCQ}
\newblock
\APACjournalVolNumPages{Lifetime data analysis}{6}{3}{187--205}.
\PrintBackRefs{\CurrentBib}

\bibitem [\protect \citeauthoryear {%
Arjas%
\ \BBA {} Saarela%
}{%
Arjas%
\ \BBA {} Saarela%
}{%
{\protect \APACyear {2010}}%
}]{%
arjas2010}
\APACinsertmetastar {%
arjas2010}%
\begin{APACrefauthors}%
Arjas, E.%
\BCBT {}\ \BBA {} Saarela, O.%
\end{APACrefauthors}%
\unskip\
\newblock
\APACrefYearMonthDay{2010}{}{}.
\newblock
{\BBOQ}\APACrefatitle {Optimal dynamic regimes: Presenting a case for
  predictive inference} {Optimal dynamic regimes: Presenting a case for
  predictive inference}.{\BBCQ}
\newblock
\APACjournalVolNumPages{The international journal of
  biostatistics}{6}{2}{1--21}.
\PrintBackRefs{\CurrentBib}

\bibitem [\protect \citeauthoryear {%
Chakraborty%
\ \BBA {} Moodie%
}{%
Chakraborty%
\ \BBA {} Moodie%
}{%
{\protect \APACyear {2013}}%
}]{%
chakraborty2013statistical}
\APACinsertmetastar {%
chakraborty2013statistical}%
\begin{APACrefauthors}%
Chakraborty, B.%
\BCBT {}\ \BBA {} Moodie, E\BPBI E.%
\end{APACrefauthors}%
\unskip\
\newblock
\APACrefYear{2013}.
\newblock
\APACrefbtitle {Statistical Methods for Dynamic Treatment Regimes} {Statistical
  methods for dynamic treatment regimes}.
\newblock
\APACaddressPublisher{}{Springer}.
\PrintBackRefs{\CurrentBib}

\bibitem [\protect \citeauthoryear {%
Jones%
}{%
Jones%
}{%
{\protect \APACyear {2010}}%
}]{%
preg}
\APACinsertmetastar {%
preg}%
\begin{APACrefauthors}%
Jones, H.%
\end{APACrefauthors}%
\unskip\
\newblock
\APACrefYearMonthDay{2010}{}{}.
\newblock
\APACrefbtitle {Reinforcement-Based Treatment for Pregnant Drug Abusers (HOME
  II).} {Reinforcement-based treatment for pregnant drug abusers (home ii).}
\newblock
\APACrefnote{\url{http://clinicaltrials.gov/ct2/show/NCT01177982}}
\PrintBackRefs{\CurrentBib}

\bibitem [\protect \citeauthoryear {%
Koo%
, Lee%
, Kim%
\BCBL {}\ \BBA {} Park%
}{%
Koo%
\ \protect \BOthers {.}}{%
{\protect \APACyear {2008}}%
}]{%
koo2008bahadur}
\APACinsertmetastar {%
koo2008bahadur}%
\begin{APACrefauthors}%
Koo, J\BHBI Y.%
, Lee, Y.%
, Kim, Y.%
\BCBL {}\ \BBA {} Park, C.%
\end{APACrefauthors}%
\unskip\
\newblock
\APACrefYearMonthDay{2008}{}{}.
\newblock
{\BBOQ}\APACrefatitle {A bahadur representation of the linear support vector
  machine} {A bahadur representation of the linear support vector
  machine}.{\BBCQ}
\newblock
\APACjournalVolNumPages{The Journal of Machine Learning
  Research}{9}{}{1343--1368}.
\PrintBackRefs{\CurrentBib}

\bibitem [\protect \citeauthoryear {%
Lavori%
\ \BBA {} Dawson%
}{%
Lavori%
\ \BBA {} Dawson%
}{%
{\protect \APACyear {2000}}%
}]{%
Lavori2000}
\APACinsertmetastar {%
Lavori2000}%
\begin{APACrefauthors}%
Lavori, P\BPBI W.%
\BCBT {}\ \BBA {} Dawson, R.%
\end{APACrefauthors}%
\unskip\
\newblock
\APACrefYearMonthDay{2000}{}{}.
\newblock
{\BBOQ}\APACrefatitle {A design for testing clinical strategies: biased
  adaptive within-subject randomization} {A design for testing clinical
  strategies: biased adaptive within-subject randomization}.{\BBCQ}
\newblock
\APACjournalVolNumPages{Journal of the Royal Statistical Society: Series A
  (Statistics in Society)}{163}{1}{29--38}.
\PrintBackRefs{\CurrentBib}

\bibitem [\protect \citeauthoryear {%
Lavori%
\ \BBA {} Dawson%
}{%
Lavori%
\ \BBA {} Dawson%
}{%
{\protect \APACyear {2004}}%
}]{%
lavori2004}
\APACinsertmetastar {%
lavori2004}%
\begin{APACrefauthors}%
Lavori, P\BPBI W.%
\BCBT {}\ \BBA {} Dawson, R.%
\end{APACrefauthors}%
\unskip\
\newblock
\APACrefYearMonthDay{2004}{}{}.
\newblock
{\BBOQ}\APACrefatitle {Dynamic treatment regimes: practical design
  considerations} {Dynamic treatment regimes: practical design
  considerations}.{\BBCQ}
\newblock
\APACjournalVolNumPages{Clinical trials}{1}{1}{9--20}.
\PrintBackRefs{\CurrentBib}

\bibitem [\protect \citeauthoryear {%
Luedtke%
\ \BBA {} van~der Laan%
}{%
Luedtke%
\ \BBA {} van~der Laan%
}{%
{\protect \APACyear {2014}}%
}]{%
laan2014}
\APACinsertmetastar {%
laan2014}%
\begin{APACrefauthors}%
Luedtke, A.%
\BCBT {}\ \BBA {} van~der Laan, M.%
\end{APACrefauthors}%
\unskip\
\newblock
\APACrefYearMonthDay{2014}{}{}.
\newblock
{\BBOQ}\APACrefatitle {Super-Learning of an Optimal Dynamic Treatment Rule}
  {Super-learning of an optimal dynamic treatment rule}.{\BBCQ}
\newblock
\APACjournalVolNumPages{U.C. Berkeley Division of Biostatistics Working Paper
  Series}{Paper 326}{}{}.
\PrintBackRefs{\CurrentBib}

\bibitem [\protect \citeauthoryear {%
Lunceford%
, Davidian%
\BCBL {}\ \BBA {} Tsiatis%
}{%
Lunceford%
\ \protect \BOthers {.}}{%
{\protect \APACyear {2002}}%
}]{%
Lunceford2002}
\APACinsertmetastar {%
Lunceford2002}%
\begin{APACrefauthors}%
Lunceford, J\BPBI K.%
, Davidian, M.%
\BCBL {}\ \BBA {} Tsiatis, A\BPBI A.%
\end{APACrefauthors}%
\unskip\
\newblock
\APACrefYearMonthDay{2002}{}{}.
\newblock
{\BBOQ}\APACrefatitle {Estimation of Survival Distributions of Treatment
  Policies in Two-Stage Randomization Designs in Clinical Trials} {Estimation
  of survival distributions of treatment policies in two-stage randomization
  designs in clinical trials}.{\BBCQ}
\newblock
\APACjournalVolNumPages{Biometrics}{58}{1}{48--57}.
\PrintBackRefs{\CurrentBib}

\bibitem [\protect \citeauthoryear {%
Moodie%
, Richardson%
\BCBL {}\ \BBA {} Stephens%
}{%
Moodie%
\ \protect \BOthers {.}}{%
{\protect \APACyear {2007}}%
}]{%
moodie2007}
\APACinsertmetastar {%
moodie2007}%
\begin{APACrefauthors}%
Moodie, E\BPBI E.%
, Richardson, T\BPBI S.%
\BCBL {}\ \BBA {} Stephens, D\BPBI A.%
\end{APACrefauthors}%
\unskip\
\newblock
\APACrefYearMonthDay{2007}{}{}.
\newblock
{\BBOQ}\APACrefatitle {Demystifying optimal dynamic treatment regimes}
  {Demystifying optimal dynamic treatment regimes}.{\BBCQ}
\newblock
\APACjournalVolNumPages{Biometrics}{63}{2}{447--455}.
\PrintBackRefs{\CurrentBib}

\bibitem [\protect \citeauthoryear {%
Murphy%
}{%
Murphy%
}{%
{\protect \APACyear {2003}}%
}]{%
murphy2003}
\APACinsertmetastar {%
murphy2003}%
\begin{APACrefauthors}%
Murphy, S\BPBI A.%
\end{APACrefauthors}%
\unskip\
\newblock
\APACrefYearMonthDay{2003}{}{}.
\newblock
{\BBOQ}\APACrefatitle {Optimal dynamic treatment regimes} {Optimal dynamic
  treatment regimes}.{\BBCQ}
\newblock
\APACjournalVolNumPages{Journal of the Royal Statistical Society: Series B
  (Statistical Methodology)}{65}{2}{331--355}.
\PrintBackRefs{\CurrentBib}

\bibitem [\protect \citeauthoryear {%
Murphy%
}{%
Murphy%
}{%
{\protect \APACyear {2005}}%
}]{%
Murphy2005}
\APACinsertmetastar {%
Murphy2005}%
\begin{APACrefauthors}%
Murphy, S\BPBI A.%
\end{APACrefauthors}%
\unskip\
\newblock
\APACrefYearMonthDay{2005}{}{}.
\newblock
{\BBOQ}\APACrefatitle {An experimental design for the development of adaptive
  treatment strategies} {An experimental design for the development of adaptive
  treatment strategies}.{\BBCQ}
\newblock
\APACjournalVolNumPages{Statistics in medicine}{24}{10}{1455--1481}.
\PrintBackRefs{\CurrentBib}

\bibitem [\protect \citeauthoryear {%
Murphy%
, Oslin%
, Rush%
\BCBL {}\ \BBA {} Zhu%
}{%
Murphy%
\ \protect \BOthers {.}}{%
{\protect \APACyear {2006}}%
}]{%
murphy2007}
\APACinsertmetastar {%
murphy2007}%
\begin{APACrefauthors}%
Murphy, S\BPBI A.%
, Oslin, D\BPBI W.%
, Rush, A\BPBI J.%
\BCBL {}\ \BBA {} Zhu, J.%
\end{APACrefauthors}%
\unskip\
\newblock
\APACrefYearMonthDay{2006}{}{}.
\newblock
{\BBOQ}\APACrefatitle {Methodological challenges in constructing effective
  treatment sequences for chronic psychiatric disorders} {Methodological
  challenges in constructing effective treatment sequences for chronic
  psychiatric disorders}.{\BBCQ}
\newblock
\APACjournalVolNumPages{Neuropsychopharmacology}{32}{2}{257--262}.
\PrintBackRefs{\CurrentBib}

\bibitem [\protect \citeauthoryear {%
Murphy%
, Van Der~Laan%
\BCBL {}\ \BBA {} Robins%
}{%
Murphy%
\ \protect \BOthers {.}}{%
{\protect \APACyear {2001}}%
}]{%
murphy2001}
\APACinsertmetastar {%
murphy2001}%
\begin{APACrefauthors}%
Murphy, S\BPBI A.%
, Van Der~Laan, M\BPBI J.%
\BCBL {}\ \BBA {} Robins, J\BPBI M.%
\end{APACrefauthors}%
\unskip\
\newblock
\APACrefYearMonthDay{2001}{}{}.
\newblock
{\BBOQ}\APACrefatitle {Marginal mean models for dynamic regimes} {Marginal mean
  models for dynamic regimes}.{\BBCQ}
\newblock
\APACjournalVolNumPages{Journal of the American Statistical
  Association}{96}{456}{1410--1423}.
\PrintBackRefs{\CurrentBib}

\bibitem [\protect \citeauthoryear {%
Oslin%
}{%
Oslin%
}{%
{\protect \APACyear {2005}}%
}]{%
naltr}
\APACinsertmetastar {%
naltr}%
\begin{APACrefauthors}%
Oslin, D.%
\end{APACrefauthors}%
\unskip\
\newblock
\APACrefYearMonthDay{2005}{}{}.
\newblock
\APACrefbtitle {Managing Alcoholism in People Who Do Not Respond to Naltrexone
  (EXTEND).} {Managing alcoholism in people who do not respond to naltrexone
  (extend).}
\newblock
\APACrefnote{\url{http://clinicaltrials.gov/ct2/show/record/NCT00115037}}
\PrintBackRefs{\CurrentBib}

\bibitem [\protect \citeauthoryear {%
Pelham~Jr%
\ \BBA {} Fabiano%
}{%
Pelham~Jr%
\ \BBA {} Fabiano%
}{%
{\protect \APACyear {2008}}%
}]{%
pelham2008}
\APACinsertmetastar {%
pelham2008}%
\begin{APACrefauthors}%
Pelham~Jr, W\BPBI E.%
\BCBT {}\ \BBA {} Fabiano, G\BPBI A.%
\end{APACrefauthors}%
\unskip\
\newblock
\APACrefYearMonthDay{2008}{}{}.
\newblock
{\BBOQ}\APACrefatitle {Evidence-based psychosocial treatments for
  attention-deficit/hyperactivity disorder} {Evidence-based psychosocial
  treatments for attention-deficit/hyperactivity disorder}.{\BBCQ}
\newblock
\APACjournalVolNumPages{Journal of Clinical Child \& Adolescent
  Psychology}{37}{1}{184--214}.
\PrintBackRefs{\CurrentBib}

\bibitem [\protect \citeauthoryear {%
Qian%
\ \BBA {} Murphy%
}{%
Qian%
\ \BBA {} Murphy%
}{%
{\protect \APACyear {2011}}%
}]{%
qian2011}
\APACinsertmetastar {%
qian2011}%
\begin{APACrefauthors}%
Qian, M.%
\BCBT {}\ \BBA {} Murphy, S\BPBI A.%
\end{APACrefauthors}%
\unskip\
\newblock
\APACrefYearMonthDay{2011}{}{}.
\newblock
{\BBOQ}\APACrefatitle {Performance guarantees for individualized treatment
  rules} {Performance guarantees for individualized treatment rules}.{\BBCQ}
\newblock
\APACjournalVolNumPages{Annals of statistics}{39}{2}{1180-1210}.
\PrintBackRefs{\CurrentBib}

\bibitem [\protect \citeauthoryear {%
Robins%
}{%
Robins%
}{%
{\protect \APACyear {2004}}%
}]{%
robins2004}
\APACinsertmetastar {%
robins2004}%
\begin{APACrefauthors}%
Robins, J\BPBI M.%
\end{APACrefauthors}%
\unskip\
\newblock
\APACrefYearMonthDay{2004}{}{}.
\newblock
{\BBOQ}\APACrefatitle {Optimal structural nested models for optimal sequential
  decisions} {Optimal structural nested models for optimal sequential
  decisions}.{\BBCQ}
\newblock
\BIn{} \APACrefbtitle {Proceedings of the second seattle Symposium in
  Biostatistics} {Proceedings of the second seattle symposium in
  biostatistics}\ (\BPGS\ 189--326).
\PrintBackRefs{\CurrentBib}

\bibitem [\protect \citeauthoryear {%
Rush%
\ \protect \BOthers {.}}{%
Rush%
\ \protect \BOthers {.}}{%
{\protect \APACyear {2004}}%
}]{%
rush2004}
\APACinsertmetastar {%
rush2004}%
\begin{APACrefauthors}%
Rush, A\BPBI J.%
, Fava, M.%
, Wisniewski, S\BPBI R.%
, Lavori, P\BPBI W.%
, Trivedi, M\BPBI H.%
, Sackeim, H\BPBI a.%
\BDBL {}Niederehe, G.%
\end{APACrefauthors}%
\unskip\
\newblock
\APACrefYearMonthDay{2004}{{\APACmonth{02}}}{}.
\newblock
{\BBOQ}\APACrefatitle {{Sequenced treatment alternatives to relieve depression
  (STAR*D): rationale and design}} {{Sequenced treatment alternatives to
  relieve depression (STAR*D): rationale and design}}.{\BBCQ}
\newblock
\APACjournalVolNumPages{Controlled Clinical Trials}{25}{1}{119--142}.
\PrintBackRefs{\CurrentBib}

\bibitem [\protect \citeauthoryear {%
Rush%
\ \protect \BOthers {.}}{%
Rush%
\ \protect \BOthers {.}}{%
{\protect \APACyear {2006}}%
}]{%
rush2006}
\APACinsertmetastar {%
rush2006}%
\begin{APACrefauthors}%
Rush, A\BPBI J.%
, Trivedi, M.%
, Wisniewski, S.%
, Nierenberg, A.%
, Stewart, J.%
, Warden, D.%
\BDBL {}Lebowitz, B.%
\end{APACrefauthors}%
\unskip\
\newblock
\APACrefYearMonthDay{2006}{}{}.
\newblock
{\BBOQ}\APACrefatitle {Acute and longer-term outcomes in depressed outpatients
  requiring one or several treatment steps: a STAR* D report} {Acute and
  longer-term outcomes in depressed outpatients requiring one or several
  treatment steps: a star* d report}.{\BBCQ}
\newblock
\APACjournalVolNumPages{American Journal of Psychiatry}{163}{11}{1905--1917}.
\PrintBackRefs{\CurrentBib}

\bibitem [\protect \citeauthoryear {%
Schneider%
\ \protect \BOthers {.}}{%
Schneider%
\ \protect \BOthers {.}}{%
{\protect \APACyear {2003}}%
}]{%
schneider2003}
\APACinsertmetastar {%
schneider2003}%
\begin{APACrefauthors}%
Schneider, L\BPBI S.%
, Ismail, M\BPBI S.%
, Dagerman, K.%
, Davis, S.%
, Olin, J.%
, McManus, D.%
\BDBL {}Tariot, P\BPBI N.%
\end{APACrefauthors}%
\unskip\
\newblock
\APACrefYearMonthDay{2003}{}{}.
\newblock
{\BBOQ}\APACrefatitle {Clinical Antipsychotic Trials of Intervention
  Effectiveness (CATIE): Alzheimer's Disease Trial.} {Clinical antipsychotic
  trials of intervention effectiveness (catie): Alzheimer's disease
  trial.}{\BBCQ}
\newblock
\APACjournalVolNumPages{Schizophrenia bulletin}{29}{1}{957--970}.
\PrintBackRefs{\CurrentBib}

\bibitem [\protect \citeauthoryear {%
Teixeira-Pinto%
, Siddique%
, Gibbons%
\BCBL {}\ \BBA {} Normand%
}{%
Teixeira-Pinto%
\ \protect \BOthers {.}}{%
{\protect \APACyear {2009}}%
}]{%
teixeira2009}
\APACinsertmetastar {%
teixeira2009}%
\begin{APACrefauthors}%
Teixeira-Pinto, A.%
, Siddique, J.%
, Gibbons, R.%
\BCBL {}\ \BBA {} Normand, S\BHBI L.%
\end{APACrefauthors}%
\unskip\
\newblock
\APACrefYearMonthDay{2009}{}{}.
\newblock
{\BBOQ}\APACrefatitle {Statistical approaches to modeling multiple outcomes in
  psychiatric studies} {Statistical approaches to modeling multiple outcomes in
  psychiatric studies}.{\BBCQ}
\newblock
\APACjournalVolNumPages{Psychiatric annals}{39}{7}{729-735}.
\PrintBackRefs{\CurrentBib}

\bibitem [\protect \citeauthoryear {%
Thall%
, Sung%
\BCBL {}\ \BBA {} Estey%
}{%
Thall%
\ \protect \BOthers {.}}{%
{\protect \APACyear {2002}}%
}]{%
thall2002}
\APACinsertmetastar {%
thall2002}%
\begin{APACrefauthors}%
Thall, P\BPBI F.%
, Sung, H\BHBI G.%
\BCBL {}\ \BBA {} Estey, E\BPBI H.%
\end{APACrefauthors}%
\unskip\
\newblock
\APACrefYearMonthDay{2002}{}{}.
\newblock
{\BBOQ}\APACrefatitle {Selecting therapeutic strategies based on efficacy and
  death in multicourse clinical trials} {Selecting therapeutic strategies based
  on efficacy and death in multicourse clinical trials}.{\BBCQ}
\newblock
\APACjournalVolNumPages{Journal of the American Statistical
  Association}{97}{457}{29-39}.
\PrintBackRefs{\CurrentBib}

\bibitem [\protect \citeauthoryear {%
Thall%
\ \BBA {} Wathen%
}{%
Thall%
\ \BBA {} Wathen%
}{%
{\protect \APACyear {2005}}%
}]{%
thall2005}
\APACinsertmetastar {%
thall2005}%
\begin{APACrefauthors}%
Thall, P\BPBI F.%
\BCBT {}\ \BBA {} Wathen, J\BPBI K.%
\end{APACrefauthors}%
\unskip\
\newblock
\APACrefYearMonthDay{2005}{}{}.
\newblock
{\BBOQ}\APACrefatitle {Covariate-adjusted adaptive randomization in a sarcoma
  trial with multi-stage treatments} {Covariate-adjusted adaptive randomization
  in a sarcoma trial with multi-stage treatments}.{\BBCQ}
\newblock
\APACjournalVolNumPages{Statistics in medicine}{24}{13}{1947--1964}.
\PrintBackRefs{\CurrentBib}

\bibitem [\protect \citeauthoryear {%
Tsiatis%
}{%
Tsiatis%
}{%
{\protect \APACyear {2006}}%
}]{%
tsiatis2006}
\APACinsertmetastar {%
tsiatis2006}%
\begin{APACrefauthors}%
Tsiatis, A\BPBI A.%
\end{APACrefauthors}%
\unskip\
\newblock
\APACrefYear{2006}.
\newblock
\APACrefbtitle {Semiparametric theory and missing data} {Semiparametric theory
  and missing data}.
\newblock
\APACaddressPublisher{}{Springer}.
\PrintBackRefs{\CurrentBib}

\bibitem [\protect \citeauthoryear {%
Wahed%
\ \BBA {} Tsiatis%
}{%
Wahed%
\ \BBA {} Tsiatis%
}{%
{\protect \APACyear {2004}}%
}]{%
wahed2004}
\APACinsertmetastar {%
wahed2004}%
\begin{APACrefauthors}%
Wahed, A\BPBI S.%
\BCBT {}\ \BBA {} Tsiatis, A\BPBI A.%
\end{APACrefauthors}%
\unskip\
\newblock
\APACrefYearMonthDay{2004}{}{}.
\newblock
{\BBOQ}\APACrefatitle {Optimal Estimator for the Survival Distribution and
  Related Quantities for Treatment Policies in Two-Stage Randomization Designs
  in Clinical Trials} {Optimal estimator for the survival distribution and
  related quantities for treatment policies in two-stage randomization designs
  in clinical trials}.{\BBCQ}
\newblock
\APACjournalVolNumPages{Biometrics}{60}{1}{124--133}.
\PrintBackRefs{\CurrentBib}

\bibitem [\protect \citeauthoryear {%
Wahed%
\ \BBA {} Tsiatis%
}{%
Wahed%
\ \BBA {} Tsiatis%
}{%
{\protect \APACyear {2006}}%
}]{%
wahed2006}
\APACinsertmetastar {%
wahed2006}%
\begin{APACrefauthors}%
Wahed, A\BPBI S.%
\BCBT {}\ \BBA {} Tsiatis, A\BPBI A.%
\end{APACrefauthors}%
\unskip\
\newblock
\APACrefYearMonthDay{2006}{}{}.
\newblock
{\BBOQ}\APACrefatitle {Semiparametric efficient estimation of survival
  distributions in two-stage randomisation designs in clinical trials with
  censored data} {Semiparametric efficient estimation of survival distributions
  in two-stage randomisation designs in clinical trials with censored
  data}.{\BBCQ}
\newblock
\APACjournalVolNumPages{Biometrika}{93}{1}{163--177}.
\PrintBackRefs{\CurrentBib}

\bibitem [\protect \citeauthoryear {%
Wathen%
\ \BBA {} Thall%
}{%
Wathen%
\ \BBA {} Thall%
}{%
{\protect \APACyear {2008}}%
}]{%
wathen2008}
\APACinsertmetastar {%
wathen2008}%
\begin{APACrefauthors}%
Wathen, J\BPBI K.%
\BCBT {}\ \BBA {} Thall, P\BPBI F.%
\end{APACrefauthors}%
\unskip\
\newblock
\APACrefYearMonthDay{2008}{}{}.
\newblock
{\BBOQ}\APACrefatitle {Bayesian adaptive model selection for optimizing group
  sequential clinical trials} {Bayesian adaptive model selection for optimizing
  group sequential clinical trials}.{\BBCQ}
\newblock
\APACjournalVolNumPages{Statistics in medicine}{27}{27}{5586--5604}.
\PrintBackRefs{\CurrentBib}

\bibitem [\protect \citeauthoryear {%
Watkins%
}{%
Watkins%
}{%
{\protect \APACyear {1989}}%
}]{%
watkins1989}
\APACinsertmetastar {%
watkins1989}%
\begin{APACrefauthors}%
Watkins, C\BPBI J\BPBI C\BPBI H.%
\end{APACrefauthors}%
\unskip\
\newblock
\APACrefYear{1989}.
\newblock
\APACrefbtitle {Learning from delayed rewards.} {Learning from delayed
  rewards.}
\newblock
\BUPhD, University of Cambridge.
\PrintBackRefs{\CurrentBib}

\bibitem [\protect \citeauthoryear {%
Zhang%
, Tsiatis%
, Laber%
\BCBL {}\ \BBA {} Davidian%
}{%
Zhang%
\ \protect \BOthers {.}}{%
{\protect \APACyear {2012}}%
}]{%
robust2012}
\APACinsertmetastar {%
robust2012}%
\begin{APACrefauthors}%
Zhang, B.%
, Tsiatis, A\BPBI A.%
, Laber, E\BPBI B.%
\BCBL {}\ \BBA {} Davidian, M.%
\end{APACrefauthors}%
\unskip\
\newblock
\APACrefYearMonthDay{2012}{}{}.
\newblock
{\BBOQ}\APACrefatitle {A robust method for estimating optimal treatment
  regimes} {A robust method for estimating optimal treatment regimes}.{\BBCQ}
\newblock
\APACjournalVolNumPages{Biometrics}{68}{4}{1010--1018}.
\PrintBackRefs{\CurrentBib}

\bibitem [\protect \citeauthoryear {%
Zhang%
, Tsiatis%
, Laber%
\BCBL {}\ \BBA {} Davidian%
}{%
Zhang%
\ \protect \BOthers {.}}{%
{\protect \APACyear {2013}}%
}]{%
robust2013}
\APACinsertmetastar {%
robust2013}%
\begin{APACrefauthors}%
Zhang, B.%
, Tsiatis, A\BPBI A.%
, Laber, E\BPBI B.%
\BCBL {}\ \BBA {} Davidian, M.%
\end{APACrefauthors}%
\unskip\
\newblock
\APACrefYearMonthDay{2013}{}{}.
\newblock
{\BBOQ}\APACrefatitle {Robust estimation of optimal dynamic treatment regimes
  for sequential treatment decisions} {Robust estimation of optimal dynamic
  treatment regimes for sequential treatment decisions}.{\BBCQ}
\newblock
\APACjournalVolNumPages{Biometrika}{100}{3}{681--694}.
\PrintBackRefs{\CurrentBib}

\bibitem [\protect \citeauthoryear {%
Zhao%
, Kosorok%
\BCBL {}\ \BBA {} Zeng%
}{%
Zhao%
\ \protect \BOthers {.}}{%
{\protect \APACyear {2009}}%
}]{%
zhao2009}
\APACinsertmetastar {%
zhao2009}%
\begin{APACrefauthors}%
Zhao, Y.%
, Kosorok, M\BPBI R.%
\BCBL {}\ \BBA {} Zeng, D.%
\end{APACrefauthors}%
\unskip\
\newblock
\APACrefYearMonthDay{2009}{}{}.
\newblock
{\BBOQ}\APACrefatitle {Reinforcement learning design for cancer clinical
  trials} {Reinforcement learning design for cancer clinical trials}.{\BBCQ}
\newblock
\APACjournalVolNumPages{Stat Med.}{28}{}{3294--3315}.
\PrintBackRefs{\CurrentBib}

\bibitem [\protect \citeauthoryear {%
Zhao%
, Zeng%
, Laber%
\BCBL {}\ \BBA {} Kosorok%
}{%
Zhao%
\ \protect \BOthers {.}}{%
{\protect \APACyear {2014}}%
}]{%
zhao2014}
\APACinsertmetastar {%
zhao2014}%
\begin{APACrefauthors}%
Zhao, Y.%
, Zeng, D.%
, Laber, E.%
\BCBL {}\ \BBA {} Kosorok, M\BPBI R.%
\end{APACrefauthors}%
\unskip\
\newblock
\APACrefYearMonthDay{2014}{}{}.
\newblock
{\BBOQ}\APACrefatitle {New Statistical Learning Methods for Estimating Optimal
  Dynamic Treatment Regimes} {New statistical learning methods for estimating
  optimal dynamic treatment regimes}.{\BBCQ}
\newblock
\APACjournalVolNumPages{Journal of the American Statistical
  Association}{}{}{DOI: 10.1080/01621459.2014.937488}.
\PrintBackRefs{\CurrentBib}

\bibitem [\protect \citeauthoryear {%
Zhao%
, Zeng%
, Rush%
\BCBL {}\ \BBA {} Kosorok%
}{%
Zhao%
\ \protect \BOthers {.}}{%
{\protect \APACyear {2012}}%
}]{%
zhao2012}
\APACinsertmetastar {%
zhao2012}%
\begin{APACrefauthors}%
Zhao, Y.%
, Zeng, D.%
, Rush, A\BPBI J.%
\BCBL {}\ \BBA {} Kosorok, M\BPBI R.%
\end{APACrefauthors}%
\unskip\
\newblock
\APACrefYearMonthDay{2012}{}{}.
\newblock
{\BBOQ}\APACrefatitle {Estimating individualized treatment rules using outcome
  weighted learning} {Estimating individualized treatment rules using outcome
  weighted learning}.{\BBCQ}
\newblock
\APACjournalVolNumPages{Journal of the American Statistical
  Association}{107}{499}{1106--1118}.
\PrintBackRefs{\CurrentBib}

\end{thebibliography}

\end{document}